\documentclass[useAMS,usenatbib]{mn2e}
\usepackage{soul}
\usepackage{amssymb}
\usepackage{amsmath}
\usepackage{epsfig}
\usepackage{soul}
\usepackage{amssymb}
\usepackage{float}
\usepackage{times}
\usepackage{graphicx}
\usepackage{multirow}
\usepackage{sidecap}
\usepackage{color}

\title[Ordered-Subsets calibration]{Radio Interferometric Calibration via Ordered-Subsets Algorithms: OS-LS and OS-SAGE calibrations}

\author[Kazemi et al. ]{ S. Kazemi$^{1}$\thanks{E-mail:
kazemi@astro.rug.nl}, S. Yatawatta$^{2}$, S. Zaroubi$^{1}$ \\ 
$^{1}$Kapteyn Astronomical Institute, University
of Groningen, P.O. Box 800, 9700 AV Groningen, the Netherlands\\
$^{2}$ASTRON, Postbus 2, 7990 AA Dwingeloo, the Netherlands }

\begin{document}
\pagerange{\pageref{firstpage}--\pageref{lastpage}} \pubyear{2012}

\voffset=-.8in
\maketitle 
\label{firstpage}

\begin{abstract}
The main objective of this work is to accelerate the Maximum-Likelihood (ML) estimation procedure in radio interferometric calibration. We introduce the OS-LS and the OS-SAGE radio interferometric calibration methods, as a combination of the Ordered-Subsets (OS) method with the Least-Squares (LS) and Space Alternating Generalized Expectation maximization (SAGE) calibration techniques, respectively. The OS algorithm speeds up the ML estimation and achieves nearly the same level of accuracy of solutions as the one obtained by the non-OS
methods. We apply the OS-LS and OS-SAGE calibration methods to simulated observations and show that these methods have a much higher convergence rate relative to the conventional LS and SAGE techniques. Moreover, the obtained results show that the OS-SAGE calibration technique has a superior performance compared to the OS-LS calibration method in the sense of achieving more accurate results while having significantly less computational cost. 
\end{abstract}
\begin{keywords}
methods: statistical, methods: numerical, techniques: interferometric
\end{keywords}
\section{Introduction}
\label{sec:intro}
Radio interferometry is the technique of combining and correlating signals from two or more separate antennas to observe the target astronomical object with a resolution determined not by the size of a single antenna but by the area covered with all the incorporated antennas. Therefore, a much better angular resolution can be achieved using radio interferometers with multiple antennas instead of single dishes \citep{A.R.1,burke}.

The main objective of designing the new generation of radio interferometers, such as the Square Kilometer Array (SKA)\footnote{http://www.skatelescope.org}, the Murchison Widefield Array (MWA)\footnote{http://www.mwatelescope.org}, the Precision Array to Probe Epoch of Reionization (PAPER)\footnote{http://astro.berkeley.edu/\~{}dbacker/eor}, the 21-cm Array (21CMA)\footnote{http://21cma.bao.ac.cn}, the Hydrogen Epoch of Reionization Array (HERA)\footnote{http://www.reionization.org}, the Long Wavelength Array (LWA)\footnote{http://lwa.unm.edu}  and the LOw Frequency ARray (LOFAR)\footnote{http://www.lofar.org}, with the ability to collect enormous amounts of data, is improving the sensitivity, resolution and frequency coverage of observations. Therefore, to deliver their scientific goals, there is a need for processing a large amount of data and for upgrading the accuracy as well as the processing time of the existing calibration techniques.  

Propagation medium and the receivers' effect in radio interferometric data are initially unknown and have to be calibrated and corrected before imaging. Self-calibration \citep{selfcal} estimates the Maximum-Likelihood (ML) estimate of the unknowns utilizing only the measurements, and due to its high accuracy, it has become the method of choice, as in this paper, for calibrating the new generation of radio synthesis arrays. 

In the presence of additive Gaussian noise, calibration is performed as a non-linear Least-Squares (LS) optimization that calculates the ML estimation using iterative gradient based methods such as Levenberg-Marquardt (LM) method \citep{A.L.1, K.L1}. However, the LS calibration suffers from a very low convergence rate because the parameters must be updated simultaneously on a complete data space. Solving for a large number of unknowns, the Jacobian computation corresponding to the applied gradient based method is considerably costly. This makes the LS calibration impractical for calibration of giant radio telescopes like SKA with thousands of receivers.  

The convergence rate and computational efficiency of calibration is significantly improved by the recently proposed Space Alternating Generalized Expectation maximization (SAGE) calibration technique \citep{S.2**************,S.K}. SAGE method \citep{J.A.1} is a specific version of the Expectation Maximization (EM) algorithm \citep{M.2} which converges even faster than the conventional EM \citep{FA}. The SAGE algorithm partitions the data space to smaller "hidden" data spaces and at every iteration, it alternates between updating parameters on some or all of them. Obtaining the ML estimate for the parameters of these small data spaces, which carry less information compared to the complete data space, provides SAGE algorithm with a superior accuracy as well as lower computational cost compared to the LS technique. Nevertheless, there is still a need for improving the speed of calibration process especially for radio synthesis arrays such as LOFAR and SKA. 

In this paper, we utilize the Ordered-subsets (OS) algorithm for accelerating the speed of calibration. The well known OS algorithm accelerates the convergence rate of iterative ML estimations and has been widely used in medical imaging \citep{OS1, OS2}. This method decomposes the objective (likelihood) function to several sub-objective functions and updates the parameters by using the gradient of one, or some, of the sub-objective functions as an approximation to the original objective function's gradient. At the initial iterations, when the parameters are far from the optimum point, these approximations are quite reasonable since the gradient is only an approximation at those stages. Thus, they can be efficient substitutions for the gradient of the original cost function and considerably accelerate the computations of the OS algorithm. However, it must be taken into account that the highest accuracy that OS methods can achieve is the same as the one which could be obtained by the conventional (non-OS) techniques. Close to the optimal solution, OS methods generally do not converge but rather become stuck at a sub-optimal limit cycle of as many points as there are sub-objective functions. Therefore, if OS method becomes globally convergent \citep{Ahn, OS1}, it maintains exactly the same accuracy of the convergent non-OS methods. 

This paper is organized as follows: In sections \ref{Calibration Data Model} and \ref{algoritms}, we present the general data model of radio interferometric calibration and the classical LS and SAGE calibration techniques. In section \ref{pro}, we introduce the OS-LS and OS-SAGE calibration techniques in order to cut down the processing time of the conventional LS and SAGE calibration methods. The computational advantages of applying the OS type calibrations instead of the conventional methods are also shown. For the ML estimations, the LM method is applied. At the end of section \ref{pro}, we show an application of OS calibration to accelerate computations when calibrating for an individual data sample. The technique is based on partitioning data over baselines and hence could also be useful in speeding up the calibration procedure of radio telescopes with a large number of receivers. In section \ref{Results}, we give results based on simulations to demonstrate the superior convergence rate of the OS calibration schemes compared to  the non-OS ones. Finally, we draw our conclusions in section \ref{Conclusions}.

The following notations are used in this paper: Bold, lowercase letters refer to column vectors, e.g., {\bf y}. Upper case bold letters refer to matrices, e.g., {\bf C}. The transpose, Hermitian transpose, and conjugation of a matrix are presented by $(.)^T$, $(.)^H$, and $(.)^*$, respectively. The matrix Kronecker product is denoted by $\otimes$. $\mathbb{R}$ is the set of Real numbers. $\operatorname{E}\{\}$ denotes the statistical expectation operator. The real and imaginary parts of complex quantities are shown by $\mathfrak{Re}$ and $\mathfrak{Im}$, respectively.

\section{Calibration Data Model}
\label{Calibration Data Model}
In this section, the general measurement equation of a polarimetric radio interferometer is presented. For some introduction to radio polarimetry and calibration the reader is referred to \citet{J.P.1} and \citet{ J.P.2}.

Consider a radio interferometer with $N$ antennas which observes $K$ uncorrelated sources. The induced voltage at antenna $p$, ${\tilde{\bf{v}}}_{pl}$, due to radiation of the $l$-th source, ${\bf{e}}_l$, is given by ${\tilde{\bf{v}}}_{pl}={\tilde{\bf{J}}}_{pl}{{\bf{e}}}_{l}$ where ${\tilde{\bf{J}}}_{pl}$  is the complex ${2\times 2}$ Jones matrix \citep{J.P.1} corresponding to the sky and instrumental corruptions of the signal. \\
The total signal obtained at antenna $p$, ${\bf v}_p$, is a linear superposition of $K$ such signals plus the antenna's thermal noise. After correcting for geometric delays and the instrumental effects, the $p$-th antenna voltage is correlated with the other $N-1$ antennas voltages. The correlated voltages $\operatorname{E}\{{\bf v}_p{\bf v}_q^H\}$, referred to as {\em {visibility}} \citep{J.P.1} of baseline $p-q$ is given by
\begin{equation}
{\bf V}_{pq}={\bf G}_p \left( \sum_{l=1}^K {{\bf{J}}}_{pl}{{\bf{C}}}_{l}{{\bf{J}}}^H_{ql} \right) {\bf G}_q^H +{\bf{N}}_{pq},\label{s2}
\end{equation}
where ${\bf N}_{pq}$ is the baseline's additive noise and ${{\bf{C}}}_{l}=\operatorname{E}\{{{\bf{e}}}_{l}{{\bf{e}}}_{l}^H\}$ is the $l$-th source {\em {coherency}} matrix \citep{bornwolf, J.P.1}. The errors common to all directions (mainly the receiver delay and amplitude errors) are given by ${\bf G}_p$ and ${\bf G}_q$. We assume that an initial calibration, at a finer time and frequency resolution, is performed to estimate ${\bf G}_p$-s  (direction independent effects). Then, the corrected data is obtained as
\begin{equation}
\widetilde{\bf V}_{pq}={\bf G}_p^{-1} {\bf V}_{pq} {\bf G}_q^{-H},
\end{equation}
where $\widetilde{\bf V}_{pq}$ are the visibilites after correction for effects common to all directions. The remaining errors are unique to a given direction, but residual errors in ${\bf G}_p$-s are also absorbed into these errors, which are denoted by ${\bf{J}}_{pl}$ in the usual notation.
 The vectorized form of corrected visibilities are given by
\begin{equation}
{\bf v}_{pq}\equiv\mbox{vec}(\widetilde{\bf V}_{pq})=\sum_{l=1}^K {\bf s}_{pql}+{\bf n}_{pq},\label{s3}
\end{equation}
where ${\bf s}_{pql}={\bf J}^*_{ql}\otimes{\bf J}_{pl}\mbox{vec}({\bf C}_{l})$ and ${\bf n}_{pq}=\mbox{vec}({\bf G}_p^{-1} {\bf N}_{pq} {\bf G}_q^{-H})$. The unknowns of the calibration problem are the real and imaginary parts of the Jones matrices complex elements
\begin{equation*}
\pmb{\theta}=[\mbox{vec}(\mathfrak{Re}\{{\bf J}_{11}\})^T\ \mbox{vec}(\mathfrak{Im}\{{\bf J}_{11}\})^T\ \mbox{vec}(\mathfrak{Re}\{{\bf J}_{12}\})^T\ldots]^T, \label{s333}
\end{equation*}
and therefore, ${\pmb{\theta}}\in \mathbb{R}^{8KN\times 1}$. 

Consider a dataset of $\tau$ time and frequency samples that form a small enough time and frequency interval over which $\pmb{\theta}$ is invariant. Stacking up the real and imaginary parts of the instrument's visibilities and noise vectors in ${\bf y}=[\mathfrak{Re}\{{\bf v}^T_{12}\}\ \mathfrak{Im}\{{\bf v}^T_{12}\}\ \mathfrak{Re}\{{\bf v}^T_{13}\}\ \ldots]^T$  and ${\bf n}=[\mathfrak{Re}\{{\bf n}^T_{12}\}\ \mathfrak{Im}\{{\bf n}^T_{12}\}\ \mathfrak{Re}\{{\bf n}^T_{13}\}\ \ldots]^T,$ respectively, the general measurement equation becomes
\begin{equation}
{\bf y}=\sum_{l=1}^K {\bf s}_l({\pmb{\theta}})+{\bf n}.\label{s4}
\end{equation}
In (\ref{s4}), ${\bf s}_l({\pmb{\theta}})=[\mathfrak{Re}\{{\bf s}^T_{12l}\}\ \mathfrak{Im}\{{\bf s}^T_{12l}\}\ \mathfrak{Re}\{{\bf s}^T_{13l}\}\ \ldots]^T$. ${\bf y},\ {\bf n},$ and ${\bf s}_l$ are vectors of size $4\tau N(N-1)$, and the noise vector ${\bf n}$ is assumed to be white Gaussian. Calibration is the ML estimation of the unknown parameter vector ${\pmb{\theta}}$ from (\ref{s4}). Note that calibration methods could also be applied to the uncorrected visibilities of (\ref{s2}) to estimate ${\bf G}_p$ and ${\bf G}_q$ errors as well. Moreover, having a large enough $N$ and small enough $K$, there will be enough constrains to solve for the $8KN$ unknown parameters of ${\pmb{\theta}}$ using the $4\tau N(N-1)$ measurements of  ${\bf y}$.

\section{The LS and SAGE Calibration Methods}\label{algoritms}
In this section, both the LS and SAGE calibration algorithms are briefly outlined. The OS scheme is applied to both methods. 

\subsection{LS calibration}\label{Normalcal}
Since the noise vector $\bf{n}$  in the calibration data model (\ref{s4}) is assumed to be white Gaussian, LS calibration method estimates the ML estimate of ${\pmb{\theta}}\in \mathbb{R}^{8KN\times 1}$ by minimizing the sum of squared errors:
\begin{equation}
\begin{array}{c}
\widehat{{\pmb{\theta}}}=\mbox{arg}\ \mbox{min}\ ||{\bf y}-\sum_{l=1}^K {\bf s}_l({\pmb{\theta}})||^2.\\
{\pmb{\theta}}\quad\quad\quad\quad\quad\quad\quad\label{s5}
\vspace*{-2mm}\end{array}
\end{equation}
Gradient-based optimization techniques are used for solving (\ref{s5}). Among those, the LM method \citep{A.L.1, K.L1} is one of the most robust in the sense that it mostly  converges to a global optimum. Defining the cost function $\phi({\pmb{\theta}})={\bf y}-\sum_{l=1}^K {\bf s}_l({\pmb{\theta}})$, where $\phi({\pmb{\theta}})\in\mathbb{R}^{4\tau N(N-1)\times 1}$, and initializing the starting point $\widehat{\pmb{\theta}}^1$, the LS calibration method via LM algorithm is outlined as follows: \\
{\bf for}\hspace*{2mm} every iteration $k=1,2,\ldots$ until an upper limit or convergence of $\widehat{\pmb{\theta}}^k$,
\begin{description}
\item{Calculate $\widehat{\pmb{\theta}}^{k+1}\in \mathbb{R}^{8KN\times 1}$ using LM algorithm as
\begin{equation}
\widehat{\pmb{\theta}}^{k+1}=\widehat{\pmb{\theta}}^k-({\pmb{\bigtriangledown}}^T_{\pmb{\theta}} {\pmb{\bigtriangledown}}_{\pmb{\theta}} +\lambda {\bf H})^{-1}{\pmb{\bigtriangledown}}^T_{\pmb{\theta}} \phi({\pmb{\theta}})|_{\widehat{\pmb{\theta}}^k}.\label{s6}
\end{equation}
}
\end{description}
{\bf endfor}

In (\ref{s6}), ${\pmb{\bigtriangledown}}_{\pmb{\theta}}=\frac{\partial }{\partial {\pmb{\theta}}} \phi({\pmb{\theta}})$, $\lambda$ is the damping factor \citep{Dampingterm}, and ${\bf H}=\mbox{diag}({\pmb{\bigtriangledown}}^T_{\pmb{\theta}} {\pmb{\bigtriangledown}}_{\pmb{\theta}})$  is the diagonal of the Hessian matrix. The sizes of the Jacobian ${\pmb{\bigtriangledown}}_{{\pmb{\theta}}}$ and the linear system solved in (\ref{s6}) are $4\tau N(N-1)\times 8KN$ and $8KN$, respectively.  Consequently, the cost of computing ${\pmb{\bigtriangledown}}^T_{{\pmb{\theta}}} {\pmb{\bigtriangledown}}_{{\pmb{\theta}}}$ is $\mathcal{O}((8KN)^2\times4\tau N(N-1))$. Therefore, since at every iteration all the $8KN$ parameters of ${\pmb{\theta}}$ are simultaneously updated, LS calibration has a very low speed of convergence. Furthermore, estimating a large number of unknowns, the Jacobian computation also becomes considerably costly. 
 
\subsection{SAGE calibration}\label{SAGE}
In the case of solving for multiple sources in the sky, the SAGE calibration algorithm \citep{S.K, S.2**************} has a significantly improved computational cost and convergence rate compared to the LS calibration. The key point is that, in general, the SAGE algorithm \citep{J.A.1} partitions the complete data space to smaller "hidden" data spaces and estimates parameters in them rather than in the complete data space. Applying the SAGE algorithm to the calibration problem, the contribution of every $l$-th source in the observation is assumed to depend only on a subset of parameters, ${\pmb{\theta}}_l \in \mathbb{R}^{8N\times 1}$. Therefore, the parameter vector ${\pmb{\theta}}\in \mathbb{R}^{8KN\times 1}$ could be partitioned for different directions (sources) in the sky as
\begin{equation*}
 {\pmb{\theta}}=[{\pmb{\theta}}^T_1\ {\pmb{\theta}}^T_2\ldots{\pmb{\theta}}^T_K]^T.
\end{equation*}
This partitioning is justifiable when the sources are sufficiently separated from each other. Initializing a starting parameter vector $\widehat{\pmb{\theta}}^1$, where $\widehat{\pmb{\theta}}^k$ denotes the estimate of ${\pmb{\theta}}$ obtained at the $k$-th iteration, SAGE calibration algorithm is executed as follows:\\
{\bf for}\hspace*{2mm} every iteration $k=1,2,\ldots$ until an upper limit for $k$ or convergence of $\widehat{\pmb{\theta}}^k$:
\begin{description}
\item{ {\bf for}\hspace*{2mm} all or some $l\in\{1,2,\ldots,K\}$, update the $l$-th source parameters ${\pmb{\theta}}_l \in \mathbb{R}^{8N\times 1}$:
\begin{enumerate}
\item Define the hidden data space as
\begin{equation}
{\bf x}_{l}={\bf s}_l({\pmb{\theta}}_l)+{\bf n}\in\mathbb{R}^{4\tau N(N-1)\times 1}.\label{n17}
\end{equation}
Thus, the observed data ${\bf y}\in\mathbb{R}^{4\tau N(N-1)\times 1}$ is given by
\begin{equation}
{\bf y}={\bf x}_l+\sum_{\substack{z=1\\z\neq l}}^K{\bf s}_z({\pmb{\theta}}_z).\label{n18}
\vspace*{-2mm}\end{equation}
\item {\em {SAGE E Step}}: Calculate the conditional mean $\widehat{{\bf x}}_l^k=\operatorname{E}\{{\bf x}_l|{\bf y},{\widehat{\pmb{\theta}}}^k\}$ as
\begin{equation*}
\widehat{{\bf x}}_l^k={\bf s}_l(\widehat{\pmb{\theta}}_l^k)+({\bf y}-\sum_{z=1}^K{\bf s}_z(\widehat{\pmb{\theta}}_z^k))={\bf y}-\sum_{\substack{z=1\\z\neq l}}^K{\bf s}_z(\widehat{\pmb{\theta}}_z^k).\label{n19}
\vspace*{-2mm}\end{equation*} 
\item {\em {SAGE M Step}}: Estimate 
\begin{equation*}
\begin{array}{c}
\widehat{\pmb{\theta}}_l^{k+1}=\mbox{arg}\ \mbox{min}\ ||[\widehat{{\bf x}}_l^k-{\bf s}_l({\pmb{\theta}}_l)]||^2,\\
{\pmb{\theta}}_l\quad\quad\quad\quad
\end{array}\vspace*{-3mm}
\end{equation*}\\
by the LM method as
\begin{equation}
\widehat{\pmb{\theta}}_l^{k+1}=\widehat{\pmb{\theta}}_l^k-({\pmb{\bigtriangledown}}^T_{{\pmb{\theta}}_l} {\pmb{\bigtriangledown}}_{{\pmb{\theta}}_l} +\lambda {\bf H})^{-1}{\pmb{\bigtriangledown}}^T_{{\pmb{\theta}}_l} \phi({\pmb{\theta}}_l)|_{\widehat{\pmb{\theta}}_l^k},\label{nn1}
\end{equation}
where $\phi({\pmb{\theta}}_l)=[\widehat{{\bf x}}_l^k-{\bf s}_l({\pmb{\theta}}_l)]\in \mathbb{R}^{4\tau N(N-1)\times 1}$.
\end{enumerate} }
\item {\bf endfor}
\end{description}
{\bf endfor}\\

Based on the above, at every $k$-th iteration, SAGE method alternates between updating parameters of some or all the sources, $l\in\{1,2,\ldots,K\}$. Calculating the ML estimate of ${\pmb{\theta}}_l \in \mathbb{R}^{8N\times 1}$ in  (\ref{nn1}), instead of the ML estimate of all parameters ${\pmb{\theta}}\in \mathbb{R}^{8KN\times 1}$ as in (\ref{s6}), it has been proved that the SAGE algorithm benefits from an accelerated convergence rate \citep{J.A.1} compared to the LS method. The sizes of the Jacobian  ${\pmb{\bigtriangledown}}_{{\pmb{\theta}}_l}$ and the linear system solved in (\ref{nn1}) are $4\tau N(N-1)\times 8N$ and $8N$, respectively. In addition, the cost of computing ${\pmb{\bigtriangledown}}^T_{{\pmb{\theta}}_l} {\pmb{\bigtriangledown}}_{{\pmb{\theta}}_l}$ is $\mathcal{O}((8N)^2\times4\tau N(N-1))$. Thus, applying LM algorithm for estimating ${\pmb{\theta}}_l$ from (\ref{nn1}), the computational expense of the SAGE calibration is much cheaper compared to the LS calibration.

Note that in the SAGE calibration, instead of partitioning the parameters of the individual sources, one could also make partitions including more than a single source sharing common parameters \citep{S.K.3}. This is more efficient when some sources have a small angular separation from each other in the sky and hence share some parameters.

\section{The OS-LS and OS-SAGE Calibration Methods}\label{pro}
In this section, OS-LS \citep{LL.} and OS-SAGE \citep{H.Z.} calibration algorithms, combinations of Ordered-Subsets (OS) algorithm with LS and SAGE calibration methods, are introduced to speed up the  conventional LS and SAGE calibration procedures.

Ordered-Subsets (OS) algorithm is applied to those optimization problems with a cost function that can be expressed as a sum of several other cost functions for accelerating the convergence rate. The solutions obtained by the OS method attain almost the same accuracy as those obtained by the non-OS optimization methods in a fraction of the time \citep{OS1}. The key idea is to consider the Jacobian of one, or some, sub-cost functions as an approximate gradient of the original cost function. These approximations are quite reasonable when one is far from the optimal point, and provide OS method with a very fast convergence rate. However, at later iterations and when the parameters are close to the global optimum, the approximations restrict the OS method to a sub-optimal limit cycle (the optima of the individual sub-observations which are processed in OS iterations). Therefore, the OS method does not converge globally \citep{Ahn}. 

Denote the visibility vectors of the $\tau$ time and frequency samples that have the fixed gain errors $\pmb{\theta}\in \mathbb{R}^{8KN\times 1}$ by ${\bf y}_1, {\bf y}_2, \ldots, {\bf y}_{\tau}$, where ${\bf y}_t\in\mathbb{R}^{4N(N-1)\times 1}$, for $t\in\{1,2,\ldots,\tau\}$. Since the noise is statistically independent, calibration problem could be restated as 
\begin{eqnarray}
\widehat{{\pmb{\theta}}}=\mbox{arg}\ \mbox{max}\prod_{t=1}^{\tau}f_t({{\bf y}_t};{\pmb{\theta}})=\mbox{arg}\ \mbox{max}\sum_{t=1}^{\tau} \mathcal{L}_t({\pmb{\theta}}|{\bf y_t}),\\[-3mm]
{\pmb{\theta}}\quad\quad\quad\quad\quad\quad\quad\quad\quad{\pmb{\theta}}\quad\quad\quad\quad\quad\quad\quad\nonumber\label{s55}
\end{eqnarray}
where $f_t$ and $\mathcal{L}_t$ are the probability density and the log-likelihood functions for the visibility vector ${\bf y}_t$, respectively. OS algorithm is applied for accelerating the maximization of this sum of log-likelihood functions. Supposing that the following Jacobian equivalence conditions hold
\begin{equation}
{\pmb{\bigtriangledown}}_{\pmb{\theta}} \mathcal{L}_1\cong {\pmb{\bigtriangledown}}_{\pmb{\theta}} \mathcal{L}_2 \cong \ldots \cong {\pmb{\bigtriangledown}}_{\pmb{\theta}} \mathcal{L}_{\tau},\label{mman}
\end{equation}
then the OS method sequentially updates the parameters ${\pmb{\theta}}$ for one or some visibility vectors ${\bf y}_t$ (sub-observations). The solution of every sub-observation is used as the starting point of the next sub-observation. Since each sub-cost function $\mathcal{L}_t$ involves a subset of data, ${\bf y}_t$, which is independent from the others, the method is named ``ordered subsets''. Sub-observations might be ordered for updating by some scheme that gives preferences to the data items, or, as in this work, in random. An introduction to the OS algorithm is presented by \citet{Ahn}. In the following, the OS-LS and OS-SAGE methods are outlined. Note that the size of sub-observations ${\bf y}_t$-s must be grater than or equal to the number of unknown parameters in ${\pmb{\theta}}$.

\subsection{OS-LS calibration}\label{mehm1}
In the presented OS-LS calibration, the LM method is selected as the gradient-based ML estimation algorithm of the LS calibration. Starting with an initial suggestion $\widehat{\pmb{\theta}}^1 \in \mathbb{R}^{8KN\times 1}$, OS-LS is executed as:\\
{\bf for}\hspace*{2mm} every iteration $k=1,2,\ldots$ until an upper limit or convergence of $\widehat{\pmb{\theta}}^k$, run $m$ {\em OS iterations}:
\begin{description}
\item{{\bf for}\hspace*{2mm} some or all sub-observation $\{{\bf y}_t|t=1,\ldots,m\leq \tau\}$:
\begin{description}
\item  {Select ${\pmb{\theta}}^k=\widehat{\pmb{\theta}}^t$, and calculate}
\item{
\begin{equation}
{\pmb{\theta}}^{k+1}={\pmb{\theta}}^k-({\pmb{\bigtriangledown}}^T_{\pmb{\theta}} {\pmb{\bigtriangledown}}_{\pmb{\theta}}+\lambda {\bf H})^{-1}{\pmb{\bigtriangledown}}^T_{\pmb{\theta}} \phi({\pmb{\theta}})|_{{\pmb{\theta}}^k},\label{mano}
\end{equation}}
\item{where  $\phi({\pmb{\theta}})=[{\bf y}_t-\sum_{l=1}^K {\bf s}_l({\pmb{\theta}})]\in\mathbb{R}^{4N(N-1)\times 1}$. }
\item{Select $\widehat{\pmb{\theta}}^{(t \bmod m)+1}={\pmb{\theta}}^{k+1}$ for the next sub-observation.}
\end{description}
\item {\bf endfor}}
\end{description}
{\bf endfor}
 
As given above, at every LM iteration, parameters are sequentially updated for some or all sub-observations. The sizes of the Jacobian ${\pmb{\bigtriangledown}}_{{\pmb{\theta}}}$ and the linear system solved in (\ref{mano}) are $4 N(N-1)\times 8KN$ and $8KN$, respectively. Moreover, the cost of computing ${\pmb{\bigtriangledown}}^T_{{\pmb{\theta}}}{\pmb{\bigtriangledown}}_{{\pmb{\theta}}}$ is $\mathcal{O}((8KN)^2\times4 N(N-1))$. When (\ref{mman}) holds, the Jacobian is calculated only for one, or a few, number of sub-observations per iteration and hence, the OS-LS method's convergence rate is considerably increased compared to the LS method.

\subsection{OS-SAGE calibration}\label{mehm2}
In this section, the OS-SAGE calibration method is introduced. A similar OS-SAGE technique is used for positron emission tomography (PET) by \citet{H.Z.}. 

Initializing  $\widehat{\pmb{\theta}}^1 \in \mathbb{R}^{8N\times 1}$, OS-SAGE is outlined as follows:\\
{\bf for}\hspace*{2mm} every $k=1,2,\ldots$ until an upper limit for $k$ or convergence of $\widehat{\pmb{\theta}}^k$, execute $m$ {\em OS iterations}:
\begin{description}
\item{ {\bf for}\hspace*{2mm} some or all sub-observations $\{{\bf y}_t|t=1,\ldots,m\leq \tau\}$:
\begin{description}
\item{Select ${\pmb{\theta}}^k=\widehat{\pmb{\theta}}^t$.}
\begin{description}
\item{ \hspace*{2mm}{\bf for}\hspace*{2mm} all or some $l\in\{1,2,\ldots,K\}$, update the $l$-th source\\
\hspace*{4mm} parameters ${\pmb{\theta}}_l \in \mathbb{R}^{8N\times 1}$:
\begin{enumerate}
\item Define 
\begin{equation*}
{\bf y}_t={\bf x}_l+\sum_{\substack{z=1\\z\neq l}}^K{\bf s}_z({\pmb{\theta}}_z),\quad{\bf x}_{l}={\bf s}_l({\pmb{\theta}}_l)+{\bf n}.\label{n1117}
\end{equation*}
\item {\em {SAGE E Step}}: Calculate $\widehat{{\bf x}}_l^k=\operatorname{E}\{{\bf x}_l|{\bf y}_t,{{\pmb{\theta}}}^k\}$ as
\begin{equation*}
\widehat{{\bf x}}_l^k={\bf y}_t-\sum_{\substack{z=1\\z\neq l}}^K{\bf s}_z({\pmb{\theta}}_z^k),\ {\bf y}_t\in\mathbb{R}^{4N(N-1)\times 1}.\label{n1119}
\end{equation*} 
\item {\em {SAGE M Step}}: Similar to (\ref{nn1}), estimate 
\begin{equation*}
\begin{array}{c}
{\pmb{\theta}}_l^{k+1}=\mbox{arg}\ \mbox{min}\ ||[\widehat{{\bf x}}_l^k-{\bf s}_l({\pmb{\theta}}_l)]||^2,\\
{\pmb{\theta}}_l\quad\quad\quad\quad
\end{array}\vspace*{-3mm}
\end{equation*}\\
\hspace*{2mm} using the LM method, by
\begin{equation}
{\pmb{\theta}}_l^{k+1}={\pmb{\theta}}_l^k-({\pmb{\bigtriangledown}}^T_{{\pmb{\theta}}_l} {\pmb{\bigtriangledown}}_{{\pmb{\theta}}_l} +\lambda {\bf H})^{-1}{\pmb{\bigtriangledown}}^T_{{\pmb{\theta}}_l} \phi({\pmb{\theta}}_l)|_{{\pmb{\theta}}_l^k}\label{nn111}
\end{equation}
\end{enumerate}}
\end{description}
\item {\hspace*{2mm}\bf endfor}
\item Select $\widehat{\pmb{\theta}}^{(t \bmod m)+1}={\pmb{\theta}}^{k+1}$ for the next sub-observation.
\end{description}
\item {\bf endfor}}
\end{description}
{\bf endfor}

OS method reduces the data size from $4\tau N(N-1)$ to $4N(N-1)$, since it calculates the partial gradients for sub-observations ${\bf y}_t\in\mathbb{R}^{4N(N-1)\times 1}$, $t\in\{1,2,\ldots,\tau \}$, instead of the whole observed data ${\bf y}\in\mathbb{R}^{4\tau N(N-1)\times 1}$. Thus, the size of the Jacobian ${\pmb{\bigtriangledown}}_{{\pmb{\theta}}_l}$, where $\phi({\pmb{\theta}}_l)=[\widehat{{\bf x}}_l^k-{\bf s}_l({\pmb{\theta}}_l)]\in \mathbb{R}^{4N(N-1)\times 1}$, calculated by LM method for every OS iteration of the OS-SAGE
calibration at (\ref{nn111}), is $4N(N-1)\times 8N$. The size of the linear system solved in (\ref{nn111}) is $8N$ and the cost of computing ${\pmb{\bigtriangledown}}^T_{{\pmb{\theta}}_l} {\pmb{\bigtriangledown}}_{{\pmb{\theta}}_l}$ is $\mathcal{O}((8N)^2\times4 N(N-1))$. When $m\ll \tau$, the OS-SAGE method
converges much faster than the conventional SAGE algorithm for which the Jacobian size is $4\tau N(N-1)\times 8N$. On the other hand, for every $t$-th OS iteration, the updated result of the $(t-1)$-th sub-observation is used as the starting point. Every OS-SAGE iteration includes $m$ number of SAGE iterations. Therefore, at initial iterations when (\ref{mman}) holds, OS-SAGE algorithm increases the likelihood function as equivalent to SAGE method with $m$ iterations. Thus, the convergence of OS-SAGE compared with SAGE is accelerated. 

\subsection{Partitioning the baselines}\label{bp}
So far, we have divided the data into sub-observations only based on their integration time and frequency. However, there are cases in which we need to calibrate for a single time and frequency interval. For instance, consider calibrating only for the $i$-th time and frequency interval when $1\leq i\leq \tau$. To apply OS calibration to such a case, one can define sub-observations by partitioning the data vector ${\bf y}_i$ over the instrument's baselines as,
\begin{equation*}
{\bf y}_i=[{\bf y}_{i1}^T\ {\bf y}_{i2}^T\  \ldots {\bf y}_{iB}^T]^T, \quad B\ll \frac{N(N-1)}{2}.
\end{equation*}
Then, similar to (\ref{s55}), the calibration problem becomes
\begin{eqnarray}
\widehat{{\pmb{\theta}}}=\mbox{arg}\ \mbox{max}\sum_{b=1}^{B} \mathcal{L}_b({\pmb{\theta}}|{\bf y}_{ib}),\\[-3mm]
\quad\quad\quad\quad\quad\quad\quad\quad\quad{\pmb{\theta}}\quad\quad\quad\quad\quad\quad\quad\nonumber\label{s5555}
\end{eqnarray}
for which OS methods presented by sections \ref{mehm1} and \ref{mehm2} are applicable, and where {\em OS iterations} are executed over $\{{\bf y}_{ib}|b=1,\ldots,m\leq B\}$. Utilizing such an OS calibration could also be beneficial in cutting down the computational expense of calibration of interferometers with a large number of receivers. The only points that should be taken into account are:
\begin{itemize}
\item{Every partition of data (sub-observation) ${\bf y}_{ib}$, for $b\in\{1,2,\ldots,B\}$, must have visibilities from different baselines such that the baselines cover all the receivers of the instrument (or all the parameters).}
\item{The number of visibilities of every sub-observation must be equal to, or larger than, the number of calibration unknowns,
\begin{equation}
||{\bf y}_{ib}||_1\geq8KN.
\end{equation}}
\end{itemize}

\subsection{Discussion}
\begin{figure*}
\begin{center}
$\begin{array}{cc}
{
\begin{array}{l}
\hspace*{-1cm}\epsfig{file=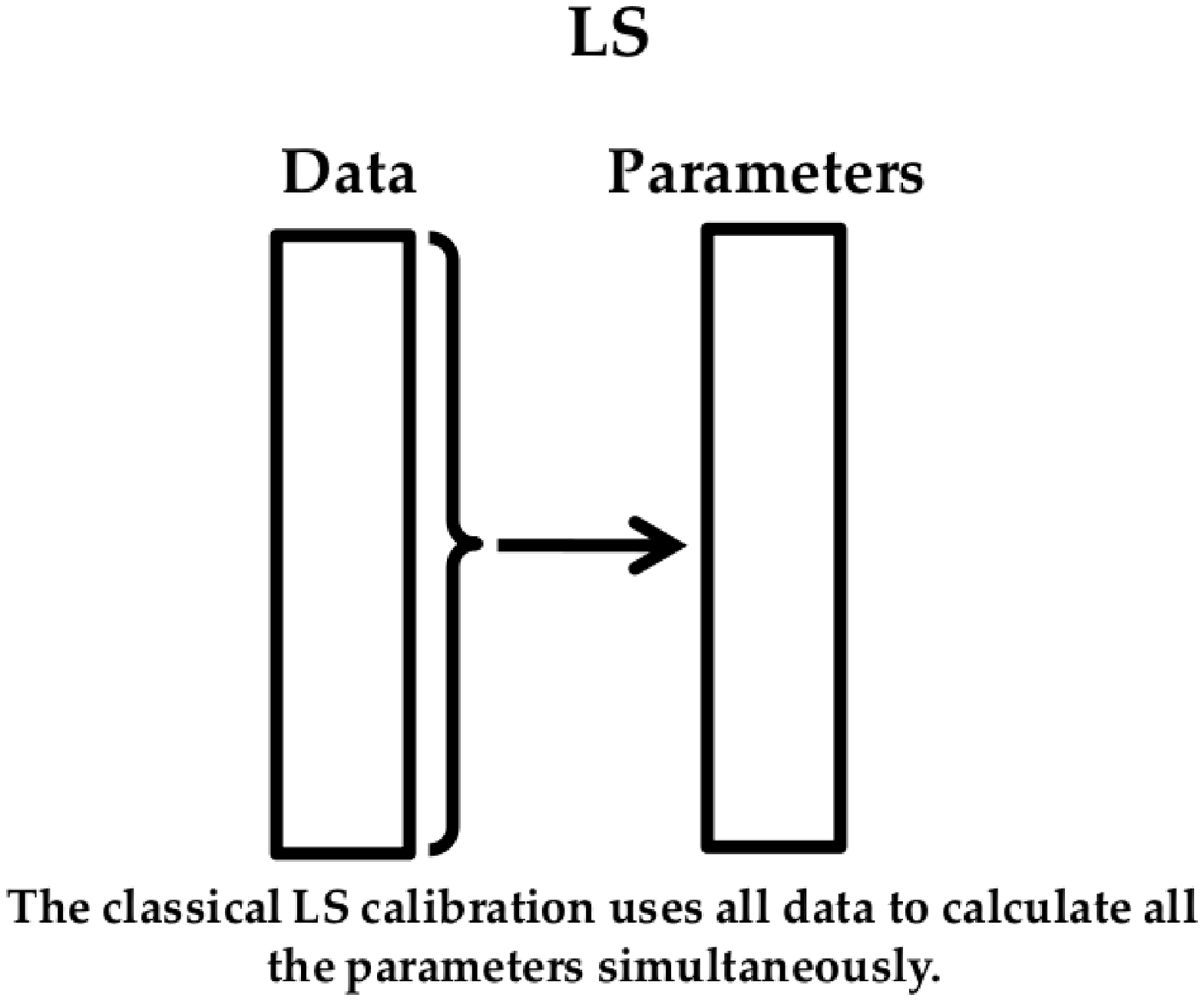, bb= 95 95 650 500,clip=,width=7cm,scale=0.4} \\[5mm]
\hspace*{-1cm}\epsfig{file=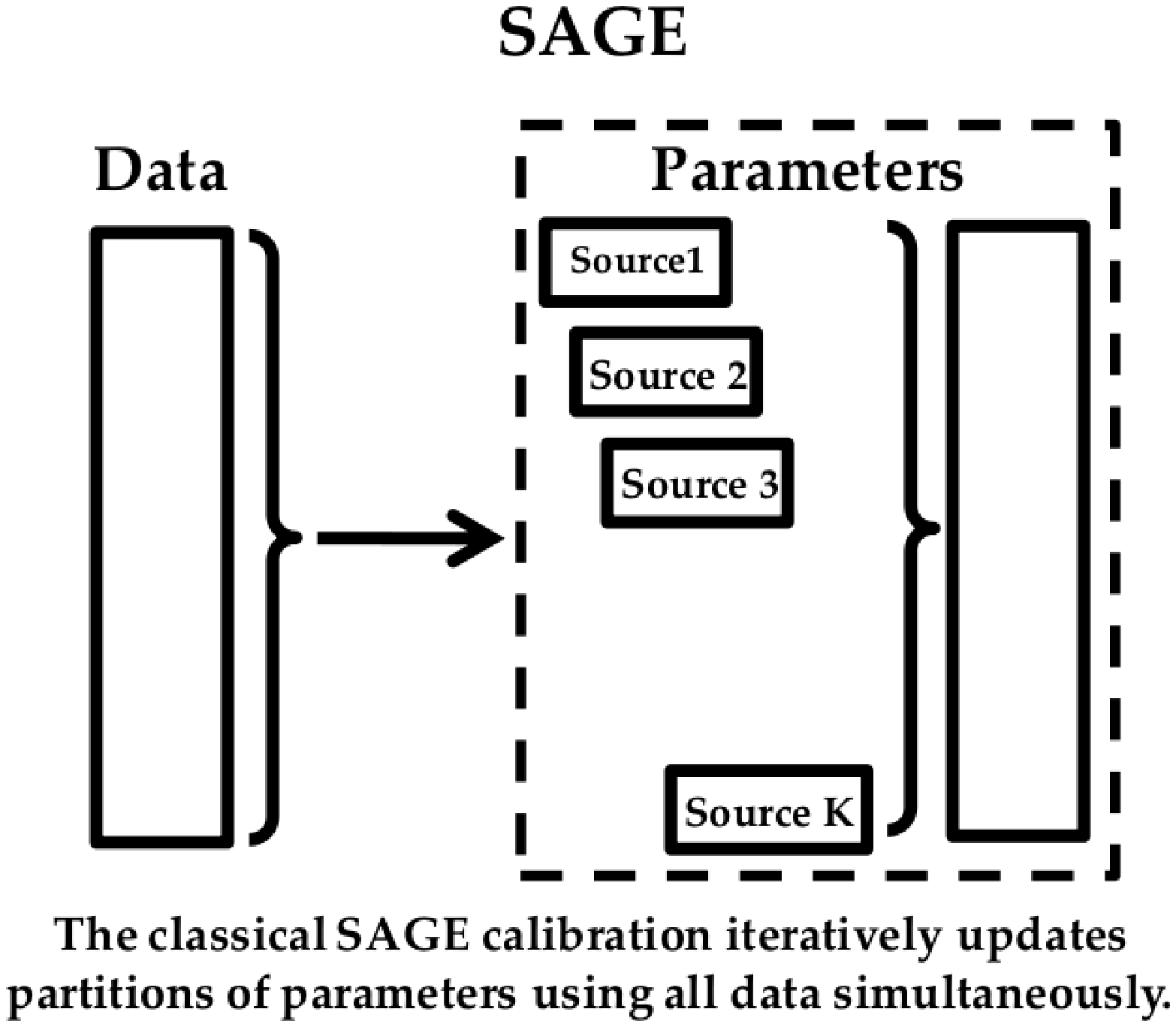, bb= 95 95 650 500,clip=,width=7cm,scale=0.4}  \\
\end{array}
}
&
{\vspace*{-5mm}\begin{array}{l}
\hspace*{1cm}\epsfig{file=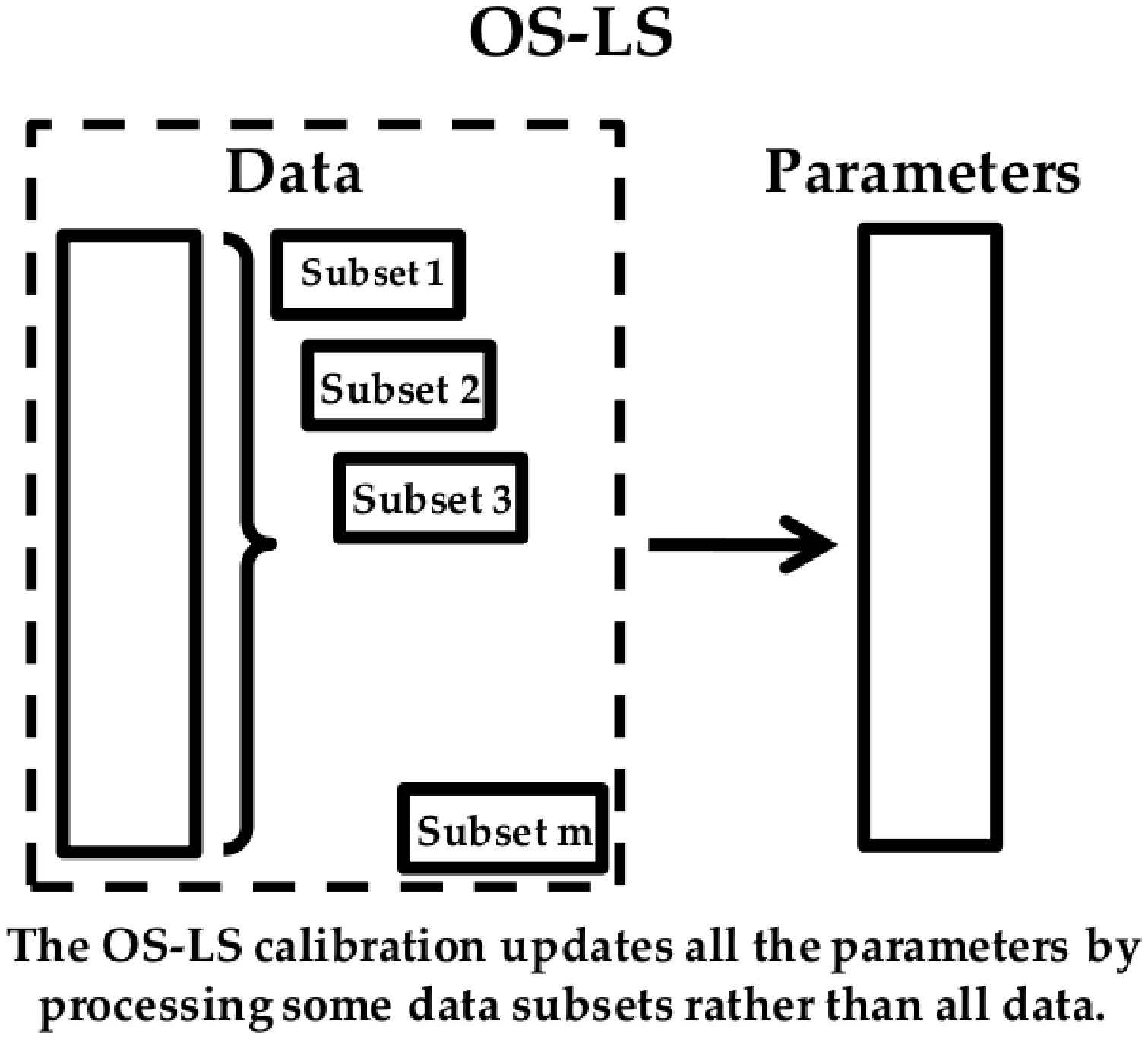, bb= 95 95 650 500,clip=,width=7cm,scale=0.4}  \\[5mm]
\hspace*{1cm}\epsfig{file=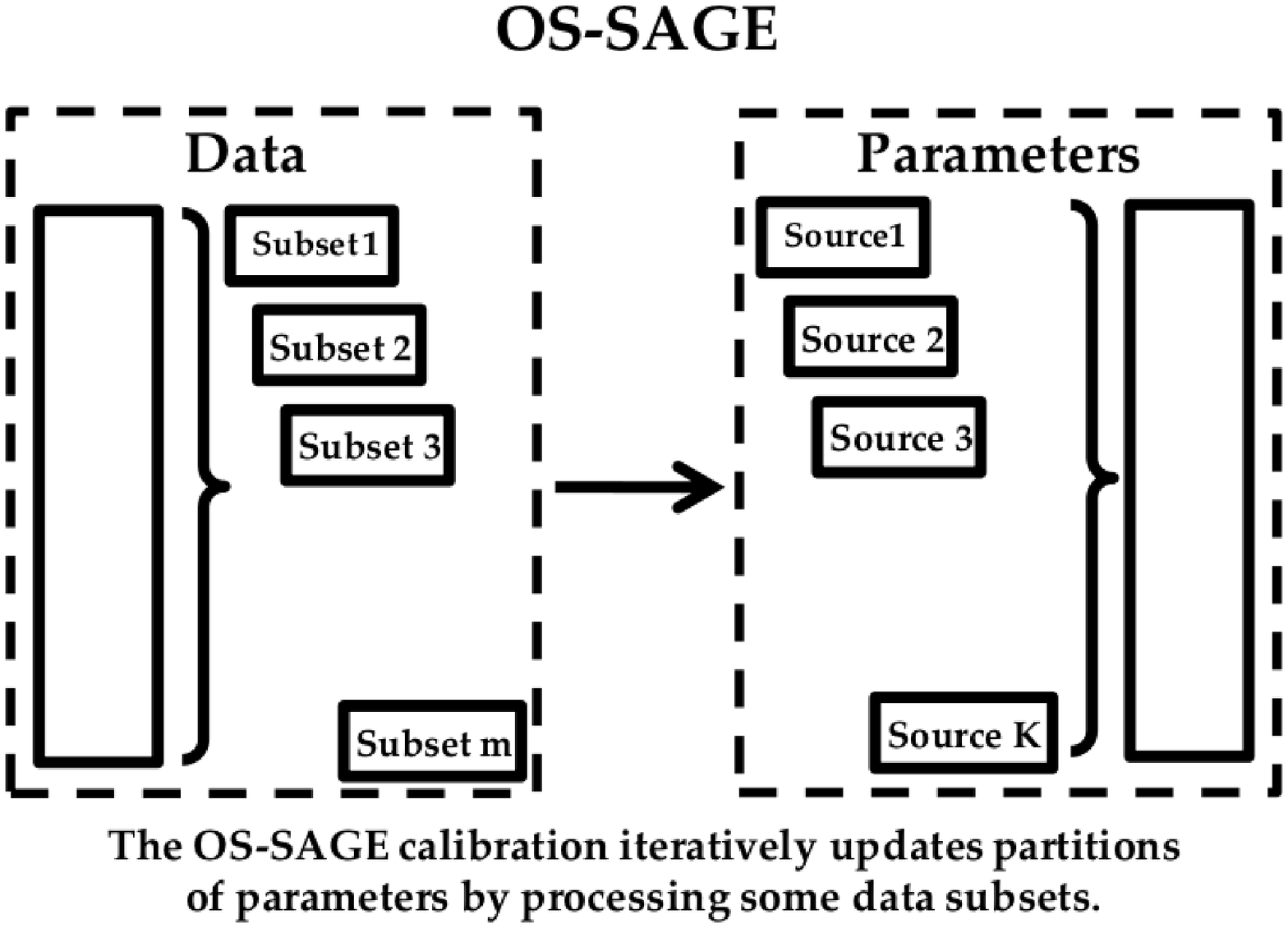, bb= 95 95 650 500,clip=,width=7cm,scale=0.4}  \\
\end{array}}
\end{array}$
\caption{Diagrams illustrating LS, SAGE, OS-LS, and OS-SAGE calibrations algorithms. }\label{modiosls}
\end{center}
\end{figure*}

To wrap up all the discussed calibration algorithms, we present a general overview in Fig. \ref{modiosls}.  Fig. \ref{modiosls}  illustrates LS, SAGE, OS-LS, and OS-SAGE calibrations algorithms.

Note that:
\begin{itemize}
\item{As it is discussed at the beginning of this section \ref{pro}, the OS algorithms do not necessarily converge. Nevertheless, there exist two major approaches in dealing with the convergence problem of the OS method: (i) using relaxation parameters (stepsizes) \citep{Ahn}. Calculating suitable relaxation parameters per every iteration is considerably costly. That makes the approach of progressively decreasing the number of sub-observations in OS method to be preferable. (ii) Reducing the number of subsets with increasing iterations until the complete dataset estimate is reconstructed \citep{OS1}.  In the OS method, one can incrementally combine some sub-observations together until there are no individual sub-observations remaining. Therefore, at the final iteration, the OS method is in fact changed to the non-OS technique which is used for the ML approximations, solving for the complete dataset. This approach guarantees global convergence as long as the non-OS ML estimation techniques (LS, SAGE, etc.) converge. However, it must be taken into account that the highest accuracy achievable by the proposed scheme is equal to any non-OS optimization methods. Modifying OS calibration in order to achieve an accuracy superior to the ones obtained by non-OS calibrations is addressed in future work. }
\item {When the Signal to Noise Ration (SNR) is poor, shifting to non-OS calibrations after running a few number of OS iterations is recommended.. Moreover, instead of running the OS method on every individual time and frequency sub-observation ${\bf y}_t$, for $t\in\{1,2,\ldots,\tau\}$, one could also apply the method to combinations of two or more sub-observations to improve the SNR. Fig. \ref{subsets} shows examples of having incrementally ordered  datasets of size two, randomly chosen datasets of the same size, and randomly chosen datasets from different sizes, from left to right, respectively. The datasets could be arranged in different orders depending on the characteristics of specific observations.  Similarly, subsets of frequency ordered sub-observations could be introduced. 
\begin{figure*}
\begin{center}
\epsfig{file=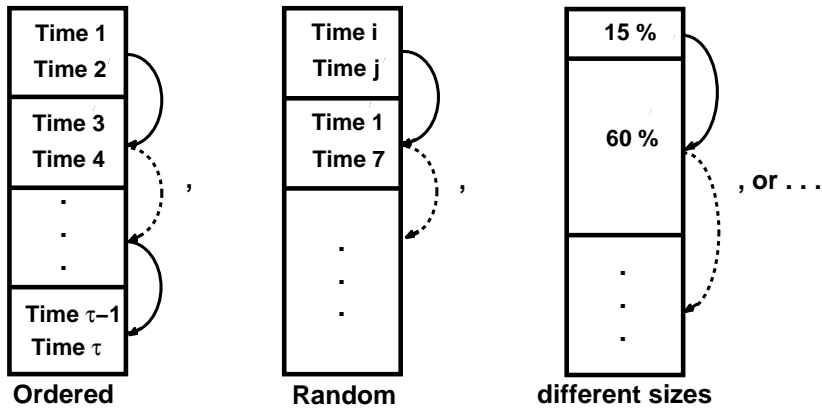, bb=  -278 102 900 683,clip=,width=12cm,scale=0.06}\\
\caption{Instead of running the OS method on every individual sub-observation, one could also apply the method to combinations of two or more sub-observations to improve the Signal to Noise Ratio (SNR). Examples of having incrementally ordered  datasets of size two, randomly chosen datasets of the same size, and randomly chosen datasets from different sizes, are shown from left to right, respectively. }\label{subsets}
\end{center}
\end{figure*}
}
\item {In the calibration data model presented by (\ref{s4}), we consider a very general form of the Jones matrices $\bf{J}$, as complex $2\times 2$ matrices, and then search for the real and imaginary parts of their elements which are collected in $\pmb{\theta}$. However, one can use a more detailed presentation of the Jones matrices in the data model, for instance, when the elements of the Jones matrices are functions of time $\zeta$ and frequency $\xi$,
\begin{equation}
{\bf J}= \left[ \begin{array}{cc}
\eta_1(\zeta,\xi) & \eta_2(\zeta,\xi)  \\
\eta_3(\zeta,\xi) & \eta_4(\zeta,\xi)  \end{array} \right].\label{gek}
\end{equation}
Then, calibration is estimation of these functions, denoted by $\eta$ in (\ref{gek}). But, this leads again to estimation of some constant parameters which define the functions. Therefore, OS calibration is also useful for such a case as well and its partitioning of data to time and frequency sub-observations would not cause any degradation of the accuracy of calibration.}
\end{itemize}

\section{Results}
\label{Results}
In this section, simulated data are used to compare the performance of LS and SAGE calibrations with OS-LS and OS-SAGE ones. Note that $n$ in this section denotes the number of iterations of the conventional LS and SAGE methods. The implementation of the calibration algorithms are done using MATLAB software. The unit of color bars of all the images are in Jansky (Jy).
\subsection{Simulations} 
\label{sim}
\begin{figure}
\centering
\epsfig{file=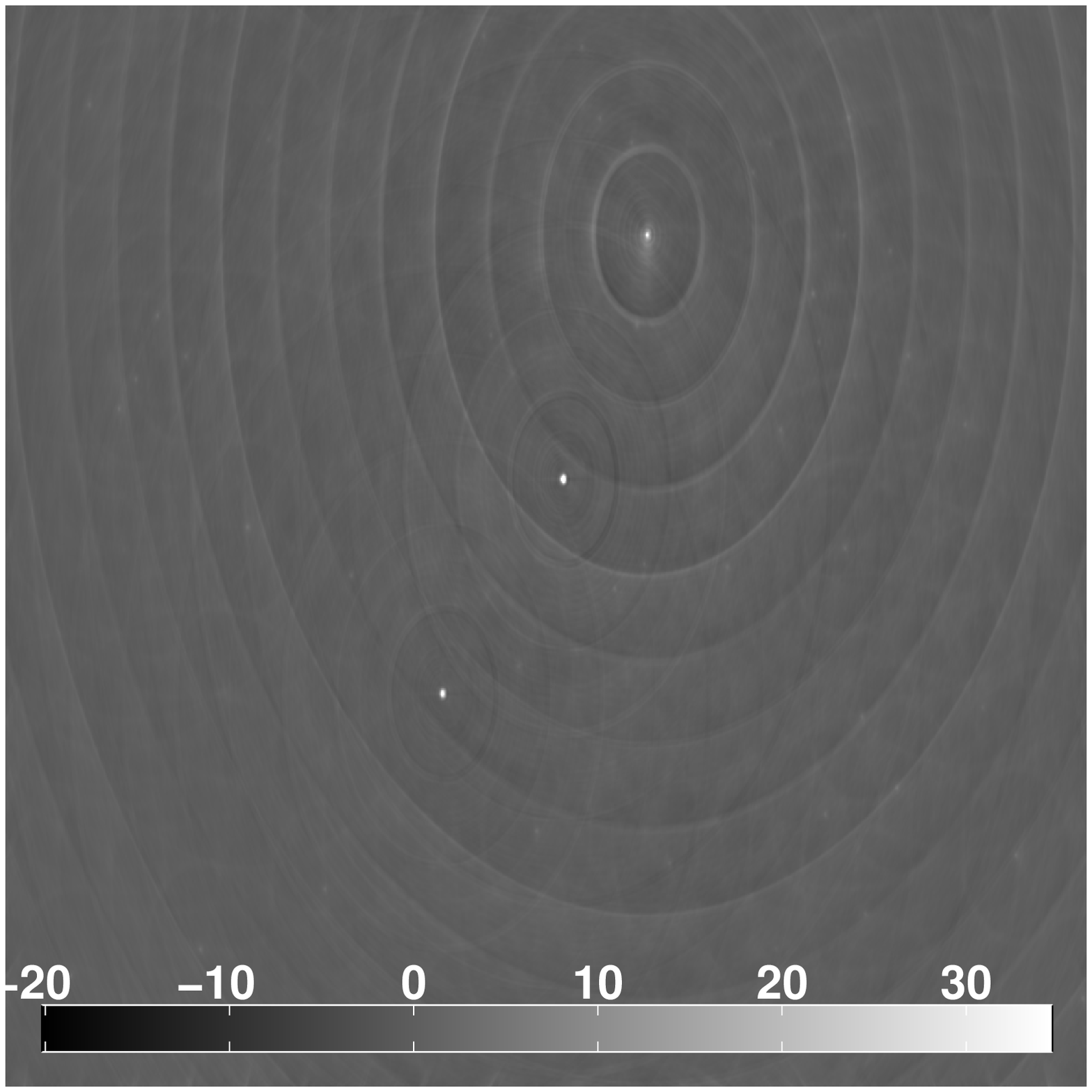, bb= 56 155 555 649,clip=,width=8.8cm,scale=0.4}
  \caption{ An $8\times 8$ degrees WSRT 12 hour simulated observation of three bright sources, with intensities 160, 107, and 108 Jy, and forty seven faint sources, with intensities below 15 Jy. The frequency is 355 MHz and the SNR is eighty.}\label{FigureA1}
\end{figure}

A 12 hour observation of Westerbork Synthesis Radio Telescope (WSRT), including 14 receivers observing a sky with 50 sources, is simulated. Three sources are very bright with intensities 160, 107, and 108 Jy, and forty seven other sources are faint with intensities below 15 Jy. The source positions are following a uniform distribution. The Jones matrices are generated as multiplications of different linear combinations of $sin$ and $cos$ functions. Their gradients vary slowly (coherence time about three minutes) as a function of time such that on a few seconds time intervals the variation could be negligible. We keep the $\mbox{SNR}=80$. The
simulated single channel image at 355 MHz is shown in Fig. \ref{FigureA1} in which the background faint sources are almost invisible. 

\begin{figure*}\vspace*{10cm}
\begin{center}$
\begin{array}{cc}
\multicolumn{2}{c}{\hspace{-6mm}\epsfig{file=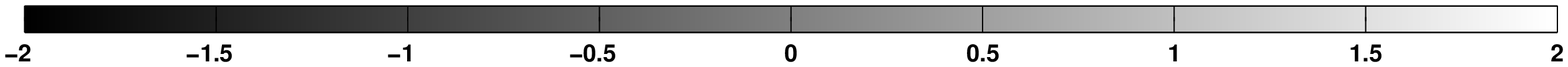, bb= -98 393 637 438, scale=0.71}}\vspace*{-10cm}\\
\hspace{-1mm}\epsfig{file=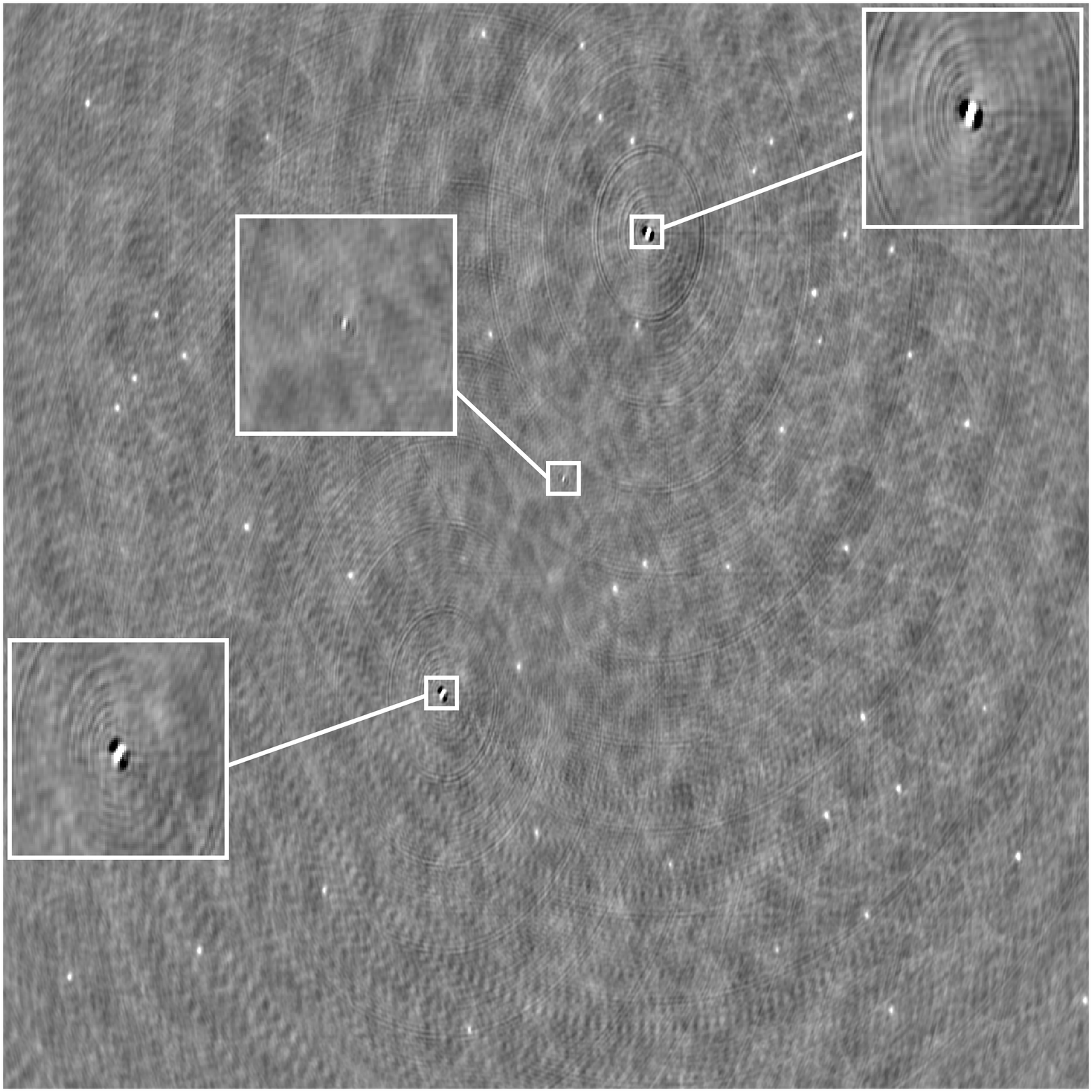, bb= -95 4 708 805,clip=,width=8.8cm,scale=0.4} &
\hspace{-3mm}\epsfig{file=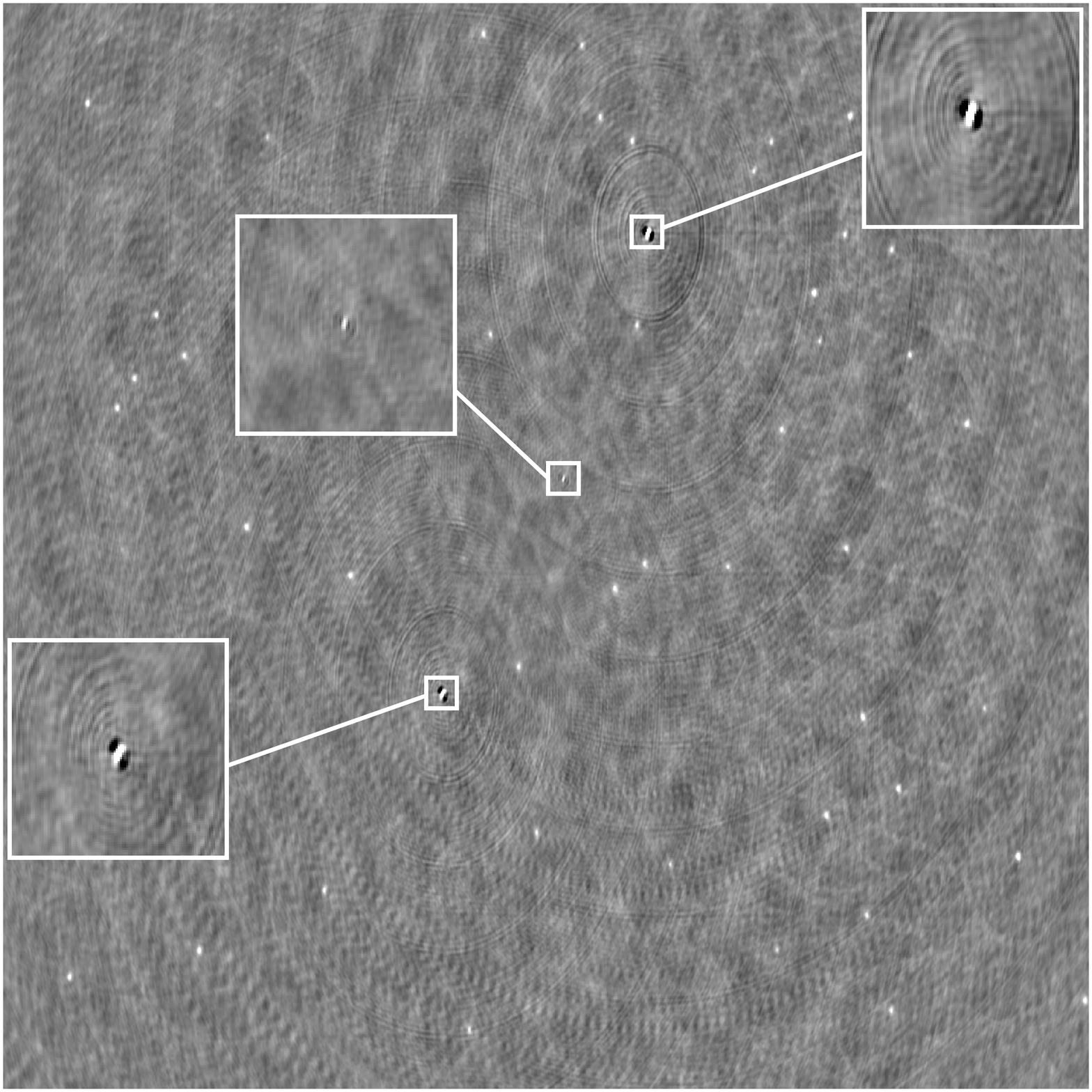, bb= -95 4 708 805,clip=,width=8.8cm,scale=0.4}\vspace*{-5mm}\\
\mbox{\color{white}\large{\bf (a)}}&\mbox{\color{white}\large{\bf (b)}}\vspace*{1cm}\\
\end{array}$
\caption{ The residual images of the LS (a) and the SAGE (b) calibrations, solving only for the three brightest sources with $n=9$ iterations. Calibrations are executed on every $\tau=10$ sub-observations simultaneously. From the three subtracted sources, the central one is the best removed (slightly underestimated) by both the SAGE and the LS calibrations. The unsolved forty seven faint sources are also visible in both (a) and (b). However, the two other bright sources are not subtracted perfectly which is expected to be improved by increasing the number of iterations. The residual noise of the SAGE algorithm is lower than of the LS method (Table \ref{table1}). This reveals the superior performance of the SAGE calibration compared to the LS calibration. }\label{FigureA2}
\end{center}
\vspace*{-5mm}
\end{figure*}

We partition the simulated data to ten seconds time intervals, $\tau=10$, including sub-observations obtained from ten individual seconds, for which the gain errors are assumed to be the same. Then, we calibrate the data partitions only for the three brightest sources via the LS and SAGE calibration methods. The residual images, obtained after $n=9$ iterations, are presented in Fig. \ref{FigureA2}. As Fig.  \ref{FigureA2} shows, among those three subtracted bright sources, the central one is the best removed (slightly underestimated) by both SAGE and the LS calibration methods. The unsolved forty seven faint sources are also visible in both residual images. But, the two other bright sources are not subtracted perfectly (overestimated in the left and right sides and underestimated in the central parts). This problematic pattern is expected to be improved by increasing the number of iterations. There is no significant difference between the residual images produced by the LS and SAGE methods in Fig. \ref{FigureA2}. However, as it is shown in Table \ref{table1}, the noise level in the  residual image of the SAGE calibration is lower than the one of the LS method. Therefore, SAGE calibration reveals a superior performance compared to the LS calibration since it achieves more accurate results with a considerably less computational complexity \citep{S.K}.  

\begin{figure*}\vspace*{19cm}
\begin{center}$
\begin{array}{cc}
\multicolumn{2}{c}{\hspace{-6mm}\epsfig{file=colormap.eps, bb= -98 393 637 438, scale=0.71}}\vspace*{-19cm}\\
\hspace{-1mm}\epsfig{file=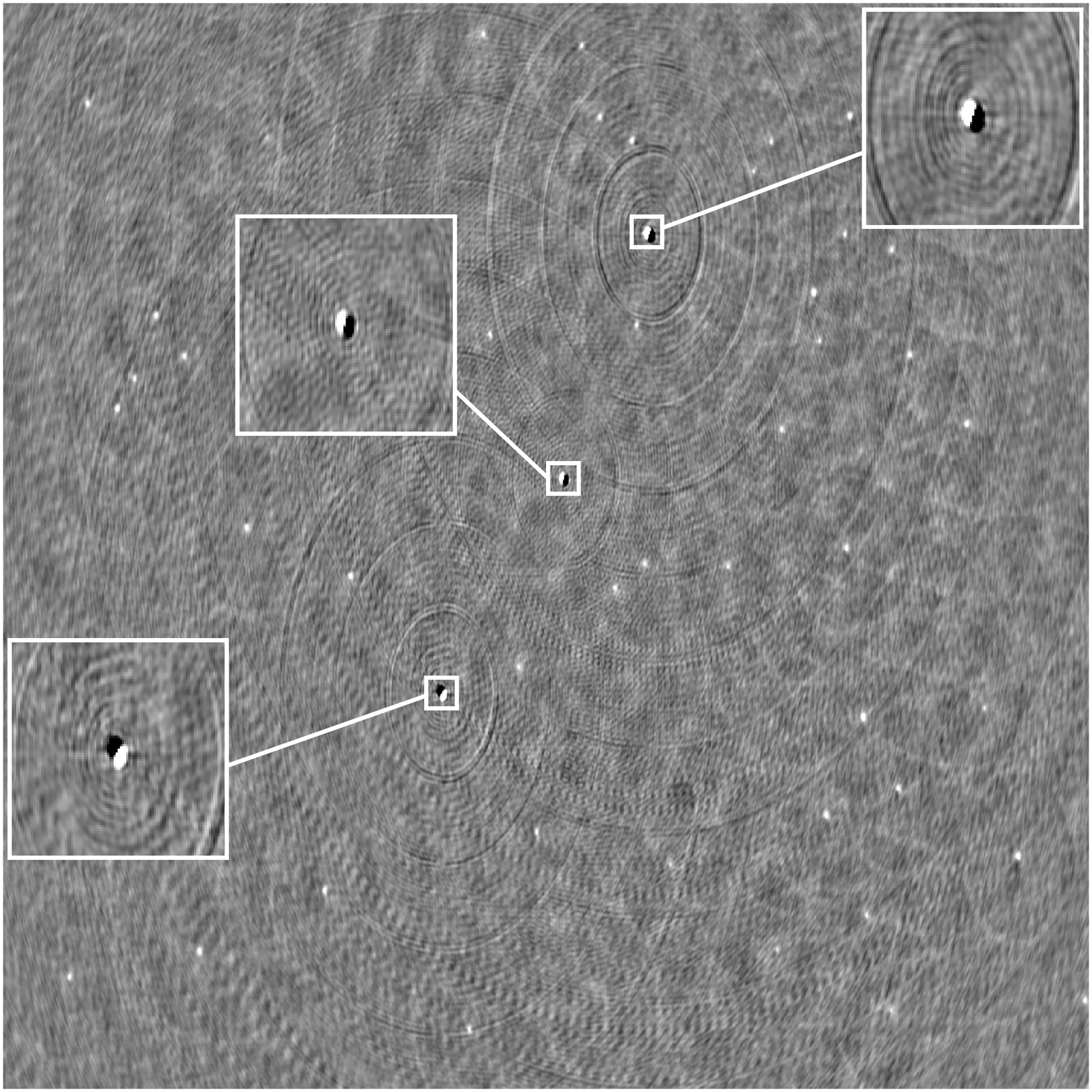, bb= -95 4 708 805,clip=,width=8.8cm,scale=0.4} &
\hspace{-3mm}\epsfig{file=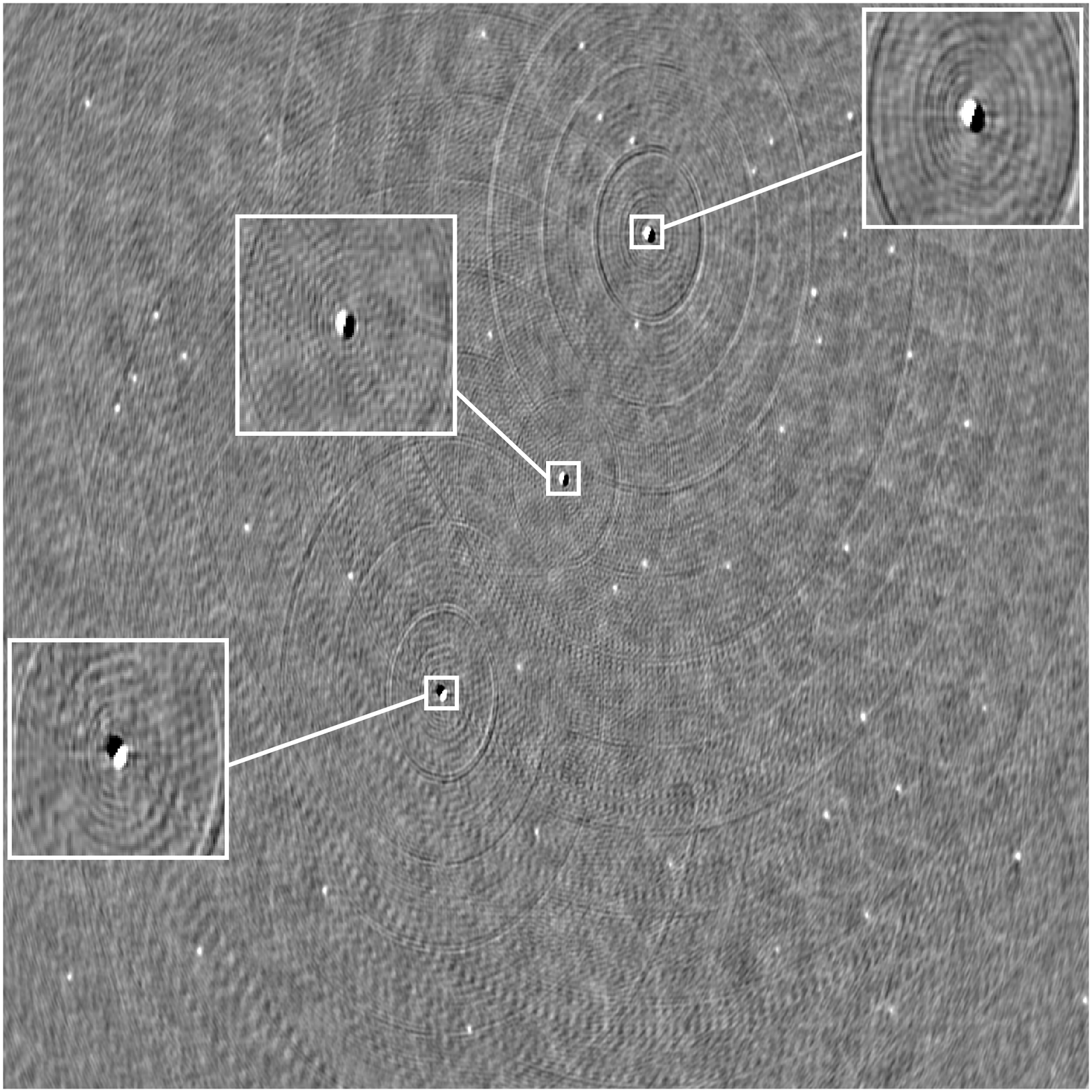, bb= -95 4 708 805,clip=,width=8.8cm,scale=0.4}\vspace*{-5mm}\\
\mbox{\color{white}\large{\bf (a)}}&\mbox{\color{white}\large{\bf (b)}}\vspace*{2mm}\\
\hspace{-1mm}\epsfig{file=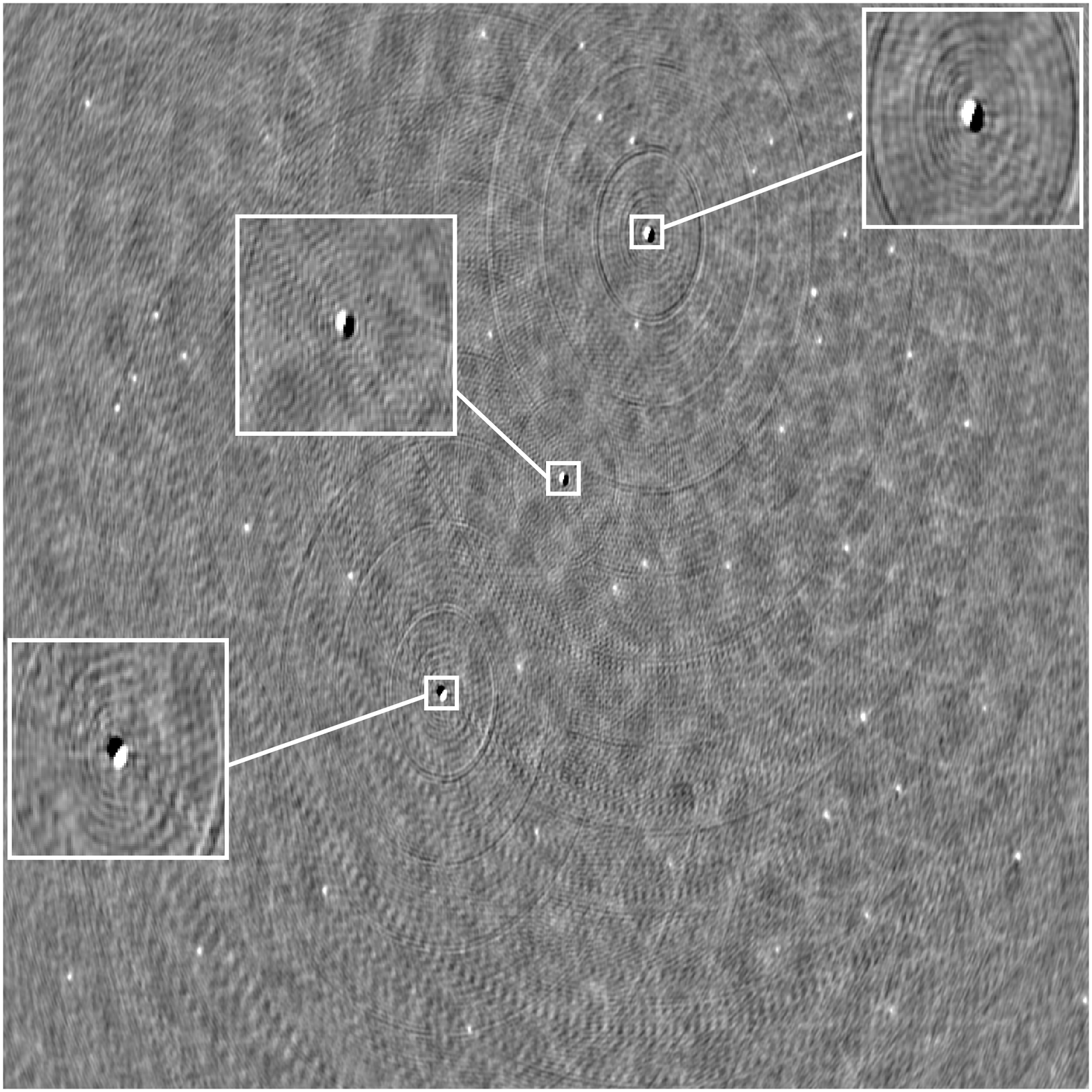, bb= -95 4 708 805,clip=,width=8.8cm,scale=0.4} &
\hspace{-3mm}\epsfig{file=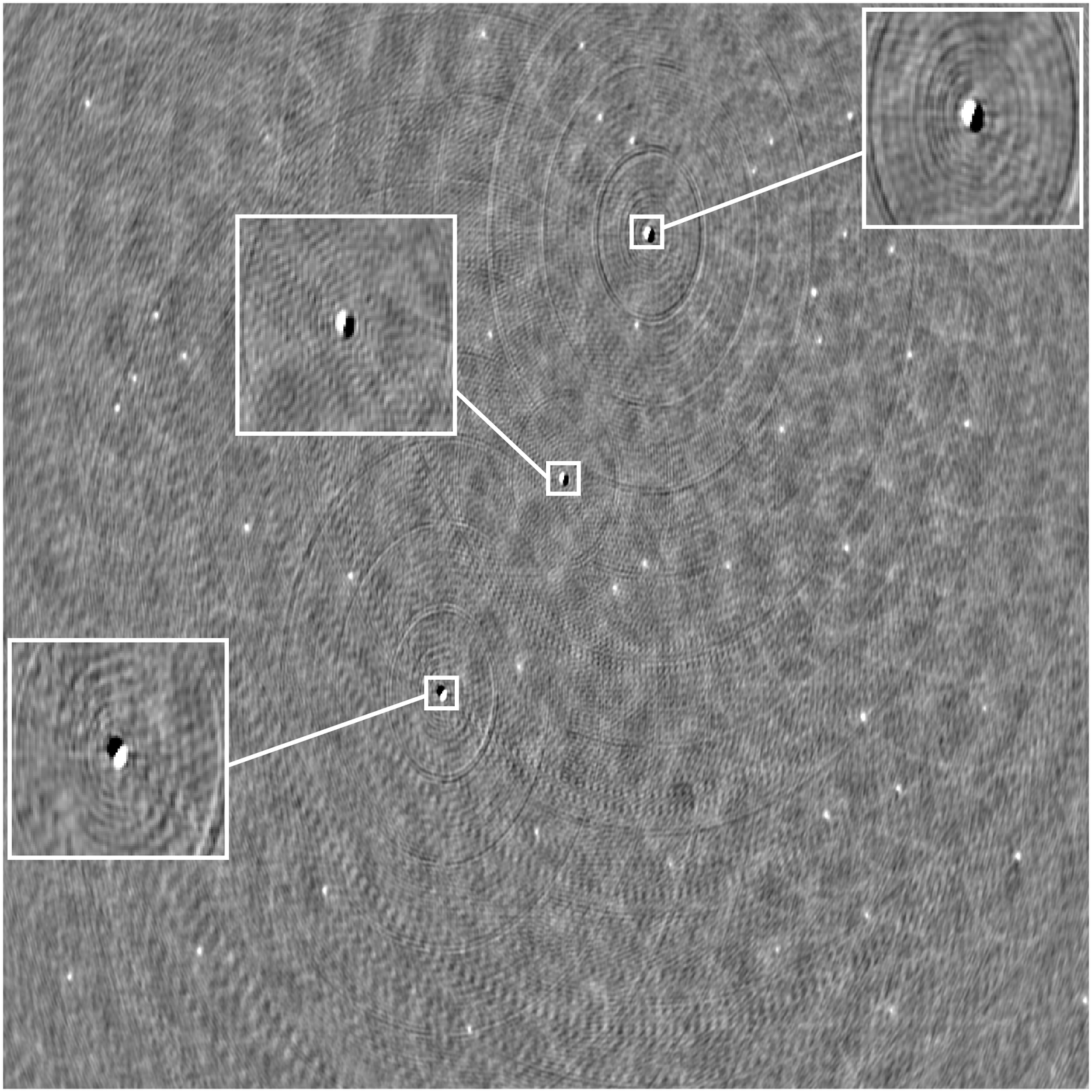, bb= -95 4 708 805,clip=,width=8.8cm,scale=0.4}\vspace*{-5mm}\\
\mbox{\color{white}\large{\bf (c)}}&\mbox{\color{white}\large{\bf (d)}}\vspace*{1cm}\\
\end{array}$
\caption{ The residual images of the OS-LS calibration for m=1 (a) and m=2 (c), and the OS-SAGE calibration for m=1 (b) and m=2 (d) OS iterations. Calibrations are executed for the three brightest sources using $n=9$ iterations. The central source is problematic in the residuals of the OS calibrations and was much better removed in the results of the conventional LS and SAGE calibrations presented in Fig. \ref{FigureA2}. Except for this source, the residual images obtained by the OS calibrations maintain the quality of the ones produced by the conventional LS and SAGE calibrations in Fig. \ref{FigureA2}. There is no visible difference between the results of m=1 and m=2 OS iterations in the images. That makes the OS calibration with m=1 OS iteration preferable in comparison with m=2 since it carries a lower computational cost. }\label{FigureA3}
\end{center}
\vspace*{-5mm}
\end{figure*}

The data is also calibrated by the OS-LS and OS-SAGE methods using $n=9$ iterations. OS iterations are executed for $m=1,2$ number of sub-observations which are randomly chosen. The residual images after subtracting the three brightest sources are presented in Fig. \ref{FigureA3}. As Fig. \ref{FigureA3} shows, the central source becomes problematic in the results of the OS calibrations and it was much better removed by the conventional LS and SAGE calibrations in Fig. \ref{FigureA2}. Except for this source, the OS calibrations have a similar quality in the residual images to the conventional LS and SAGE calibrations. The two other subtracted sources are not perfectly removed and the other forty seven faint sources are visible in the images, similar to Fig. \ref{FigureA2}. The residual images obtained for m=1 and m=2 OS iterations look almost the same. There is no  significant improvement in the residual noise level when using $m=2$ OS iterations instead of $m=1$, as it is evident in Table \ref{table1}. In this case, the OS calibration with m=1 OS iteration is preferable in comparison with m=2 since it carries a lower computational cost. 

The calibrations execution times, in minutes, and the residual noise levels, in milliJansky (mJy), are presented in Table \ref{table1}. Table \ref{table1} shows that the OS calibrations have a much faster processing speed compared to the conventional LS and SAGE calibrations. Among OS calibrations, the ones with a smaller number of OS iterations always have faster execution, as it is the case comparing the processing times for $m=1$ to $m=2$.  The fastest execution speed of the calibration method belongs to the OS calibrations with $m=1$ OS iteration.  On the other hand, the OS calibrations including a large number of OS iterations usually produce more accurate solutions since they use a higher level of information in their computations. As the results of Table \ref{table1} demonstrate, the accuracy obtained by $m=2$ number of OS iterations is slightly higher than the one achieved by $m=1$. However, the use of $m=1$ number of OS iterations is still preferred compared to $m=2$ since it has a considerably lower processing time. Note that the use of the SAGE type calibration methods are always preferred compared to the LS ones, providing more accurate results in a lower processing time.\\
\begin{table}
\begin{center}
\begin{tabular}{|l@{}|l@{}|l@{}|l@{}|l|}
\multicolumn{5}{c}{{\bf Table 1}}\\
\hline
m= number of OS iterations &\multicolumn{2}{|c|}{LS}&\multicolumn{2}{|c|}{SAGE} \\
\cline{2-5}
LS or SAGE iterations $n=9$ &Time & Noise &  Time & Noise\\ 

 & [minutes] & [mJy] &  [minutes] & [mJy]\\
\hline
OS, $m=1$                    & 41.3 & 234.2 & 9.7 & 226.1 \\
OS, $m=2$                    & 75.5 & 232.9 & 20.4 & 225.7 \\
Conventional methods & 103.9 & 180.1 & 86.3 & 179.2 \\
\hline
\end{tabular}
\end{center}
\caption{Execution times of calibration (minutes) and the standard deviation of the residual noise (mJy). The OS calibrations perform much faster than the conventional LS and SAGE calibrations. The lowest execution time of the OS results are obtained for $m=1$. On the other hand, the most accurate results are obtained for $m=2$ number of OS iterations. Moreover, SAGE type calibrations is always preferred  to the LS ones, having a higher accuracy and less computational complexity.}  
\label{table1}
\end{table}

\begin{figure*}
\centering
\hspace*{-1cm}
\includegraphics[width=20cm,height=12cm]{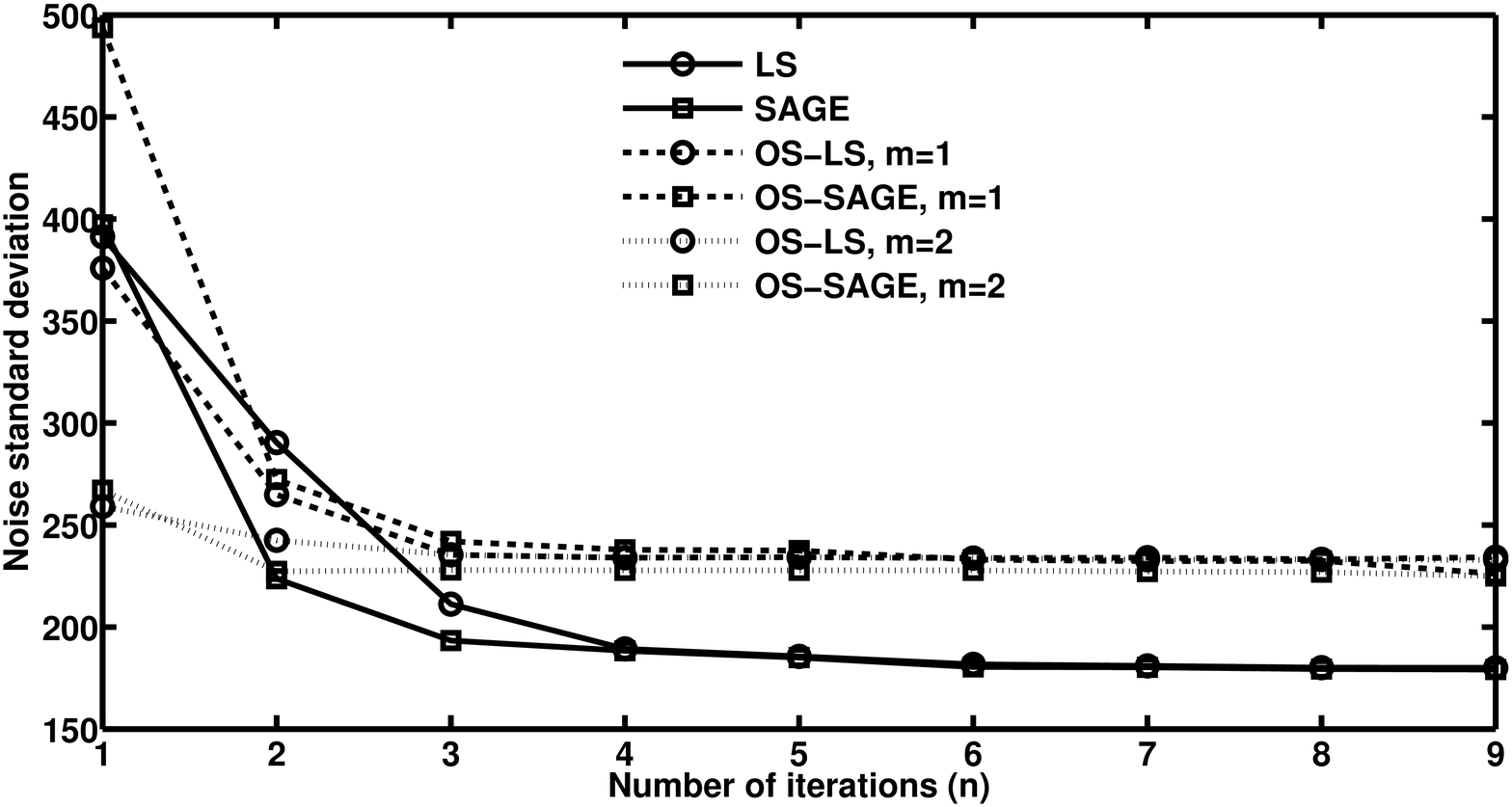}
\vspace{-0.5cm}
\caption{The residual noise standard deviations of the calibration methods in $(\mbox{mJy})$ versus their number of iterations which varies between one to nine, $n\in\{1,\ldots,9\}$. The number of OS iterations are denoted by $m$. In the plots of Fig. \ref{FigureA6}, the residual noise levels of the OS calibrations are higher than the ones obtained by the non-OS calibrations. However, it must be taken into account that these results are generated spending a comparably less computational cost compared to the classical LS and SAGE calibrations. By increasing $n$, the result of SAGE calibrations are always better than of LS calibrations. Moreover, the accuracy of OS calibrations which use $m=2$ OS iterations are superior to the one obtained by $m=1$.}
\label{FigureA6}
\end{figure*}

Fig. \ref{FigureA6} illustrates the residual noise level achieved by the calibration procedures versus the number of iterations of the LS and SAGE methods, when it varies between one to nine, $n\in\{1,\ldots,9\}$. The number of OS iterations are denoted by $m$. In the plots of Fig. \ref{FigureA6}, the residual noise levels of the OS calibrations are higher than the ones of the non-OS calibrations. However, it must be taken into account that these results are obtained by using a comparably less computational cost compared to the classical LS and SAGE calibrations. By increasing $n$, the result of SAGE calibrations are always better than the one of LS calibrations. Moreover, the accuracy of OS calibration using $m=2$ OS iterations are also always superior to the results obtained by $m=1$.

As we have seen so far in this simulation, among the OS calibrations, the ones with a smaller number of OS iterations (smaller $m$) have a lower execution time.  On the other hand, the OS calibrations including a large  number of OS iterations usually produce more accurate solutions since they use a higher level of information in their computations of the Jacobian. However, the use of a small number of OS iterations is still preferable since it is considerably faster and applying a high enough number of calibration iterations,  we would achieve the same accuracy as with large $m$. 

In this section, we also demonstrate the applicability of the OS calibration in calibrating for a single time and frequency data sample, as it is discussed in section \ref{bp}, where the data must be partitioned over the instrument's baselines. There are various ways of such a partitioning of visibilities among which we use the most efficient one for this specific simulation. 
\begin{enumerate}
\item{The first question is ``{\em what is the maximum number of partitions of data over the baselines that we can define such that the baselines of every single partition cover all the receivers of the interferometer?}''. The reason of searching the maximum is to get the highest level of information at every calibration's sub-observation later on. To answer this question, we use some well-known definitions of graph theory \citep{graph} .\\
Consider the interferometer as a complete graph of order $N$ \footnote{A complete graph of order N has N nodes and every pair of nodes are connected to each other by a unique edge.} where the receivers and the baselines are the nodes and edges of the graph, respectively. Therefore, since in this simulation $N$ is even, the answer to our question is the chromatic index of this graph which is qual to $N-1$. This means we can color the $\frac{N(N-1)}{2}$ edges of the graph by $N-1$ colors where every color is covering all the $N$ nodes and $\frac{N}{2}$ number of edges. For instance, Fig. \ref{FigureB3} shows a complete graph of order eight, colored by $8-1=7$ colors, where every color covers all the nodes by $\frac{8}{2}=4$ number of edges. We partition the visibilities based on the color of their corresponding baselines in the graph. Thus, at every partition, we have $\frac{N}{2}$ number of visibility matrices. 
\begin{figure}\hspace*{-3mm}
\centering
\epsfig{file=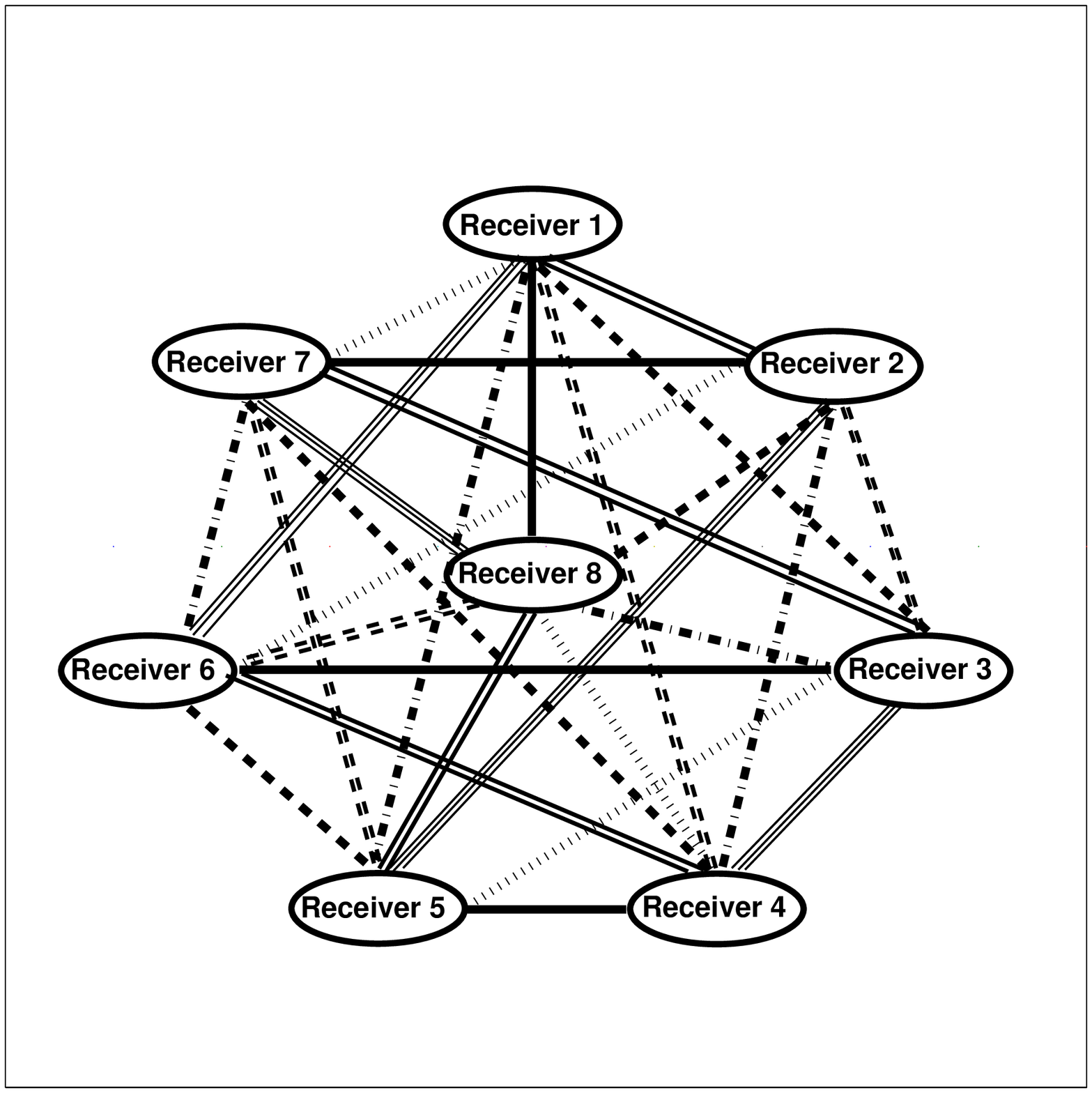, bb= 95 200 580 600,clip=,width=8.8cm,scale=0.5}
  \caption{A complete graph of order eight, colored by $8-1=7$ number of colors. Every color covers all the nodes by $\frac{8}{2}=4$ number of edges. }\label{FigureB3}
\end{figure} 
}
\item{The second question is ``{\em how many partitions should be collected at every OS calibration's sub-observation to ensure that (\ref{s4}) is not an under-determined system?'}'. Every partition has $\frac{N}{2}$ of baselines and we are trying to estimate $KN$ Jones matrices. Therefore, we must have at least $x$ partitions at every OS calibration's sub-dataset where
\begin{equation}
x \frac{N}{2} > KN.
\end{equation}
Thus,
\begin{equation}
x \geq 2K+1.
\end{equation}
}
\end{enumerate}

\begin{figure}\hspace*{-3mm}
\centering
\epsfig{file=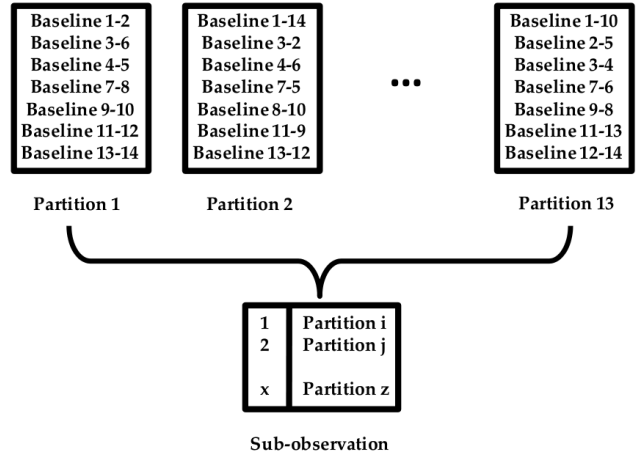, bb= 44 63 679 520,clip=,width=8.6cm,scale=0.5}
  \caption{There are $N=14$ number of receivers in WSRT and hence $N-1=13$ number of partitions over its baselines, each including  $\frac{N}{2}=7$ number of visibilities and covering all the receivers. Every OS sub-observation consists of $x\leq 7$ number of such partitions. Thus, at every sub-observation we have $x\times \frac{N}{2} $ number of visibility matrices.  }\label{FigureB5}
\end{figure}

We have $N=14$ number of receivers in WSRT. Thus, $\frac{N(N-1)}{2}=91$ number of baselines, providing $2 \times 2$ visibility matrices, at every time and frequency sample. According to (i) we can make thirteen partitions of baselines so that every partition includes $\frac{N}{2}=7$ number of visibilities covering all the receivers. Since we calibrate for $K=3$ bright sources A, B, and C, using (ii), $x \geq 7$. This means at every OS sub-observation we must collect at least seven number of those partitions. Thus, at every sub-observation we have $x\times \frac{N}{2} =49$ number of visibility matrices and that is enough for estimating $KN=42$ number of Jones matrices. Indeed better accuracy of OS calibration is expected to be obtained by increasing $x$ till $x\leq N-1$. This approach of defining sub-observations of the OS calibration is demonstrated in Fig. \ref{FigureB5}. As this figure shows, there are no overlaps between the baselines of the thirteen different partitions. Therefore, the maximum information level, achievable by using $x\times\frac{N}{2}=49$ number of visibilities, is provided for every sub-observation of the OS calibration.

\begin{figure*}\vspace*{10cm}
\begin{center}$
\begin{array}{cc}
\multicolumn{2}{c}{\hspace{-7mm}\epsfig{file=colormap.eps, bb= -98 393 637 438, scale=0.72}}\vspace*{-10cm}\\
\hspace{-2mm}\epsfig{file=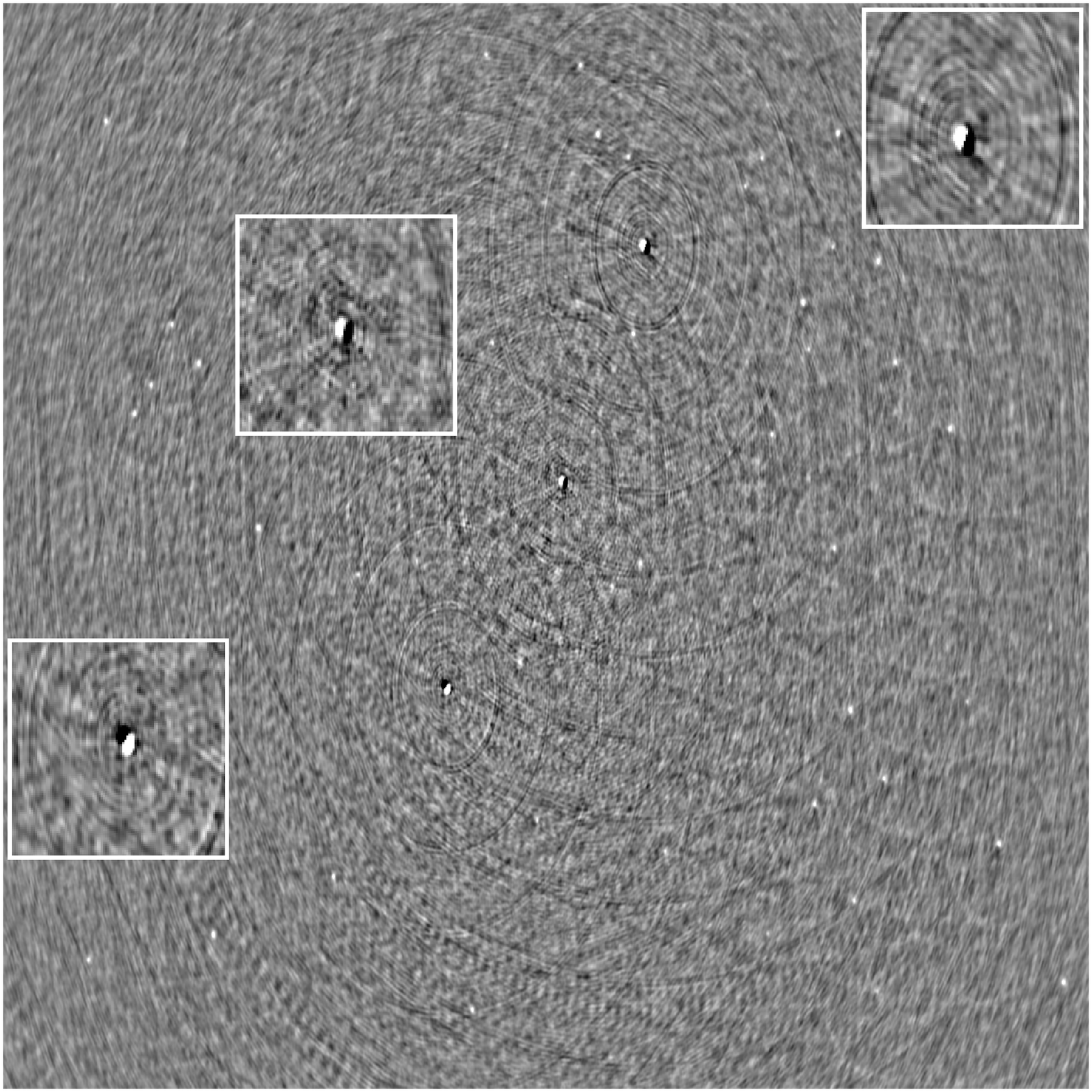, bb= -87 12 700 800,clip=,width=8.8cm,scale=0.4} &
\hspace{-3mm}\epsfig{file=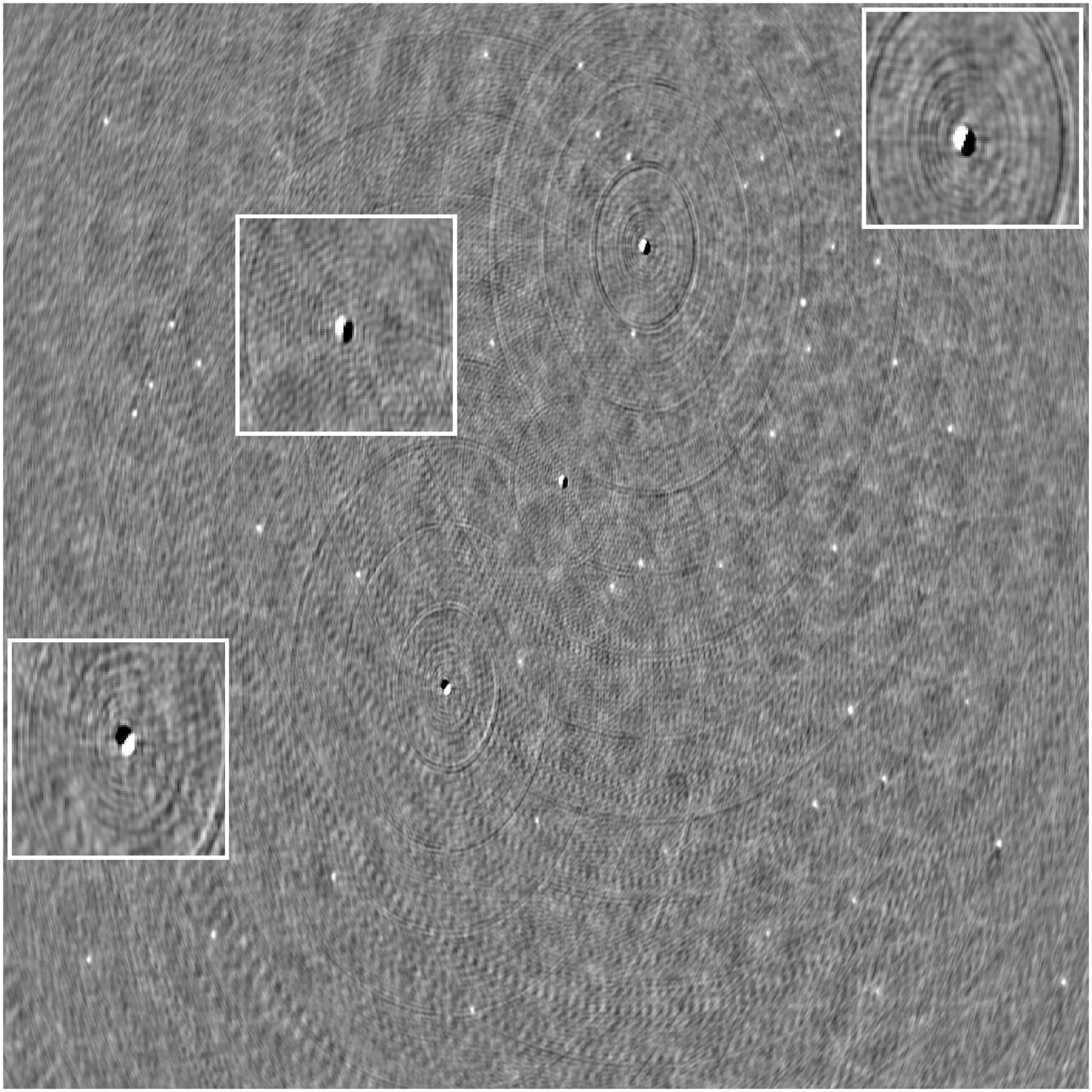, bb= -87 12 700 800,clip=,width=8.8cm,scale=0.4}\vspace*{-5mm}\\
\mbox{\color{white}\large{\bf (a)}}&\mbox{\color{white}\large{\bf (b)}}\vspace*{1cm}\\
\end{array}$
\caption{ The residual images obtained by the OS-SAGE calibration, using $m=2$ number of sub-observations at every iteration, for $x=7$ (a) and $x=10$ (b). By increasing the number of visibilities in the sub-observations from forty nine ($x=7$) to seventy ($x=10$), the calibration accuracy is highly improved. Plus, the two images have a higher residual noise and artifacts compared to the result obtained for $x=N-1=13$, which is presented by Fig. \ref{FigureA3} as image (d). This shows that better accuracy of the OS calibration is achieved when there exist a larger number of visibilities in every sub-observation. However, the calibration's processing time for $x=7$ and $x=10$ is $73.5$ and $92.8$ minutes, respectively, which is faster than the one for $x=13$ that was $108.8$ minutes (Table 1).}\label{FigureB4}
\end{center}
\end{figure*}

OS-SAGE calibration is executed, using $m=2$ number of time samples at every iteration (two number of OS iterations), for $x=7$ and $x=10$. The residual images are shown in Fig. \ref{FigureB4}. We can see that by increasing the number of visibilities in the sub-observations from forty nine ($x=7$) to seventy ($x=10$), the calibration accuracy is highly improved. We also can see that the two images of Fig. \ref{FigureB4} have a higher residual noise and artifacts compared to the result obtained for $x=N-1=13$, which is presented by Fig. \ref{FigureA3} as image (d). This shows that better accuracy of the OS calibration is achieved when the number of visibilities in every sub-observation is large. However, the calibration's processing times for $x=7$ and $x=10$ are $73.5$ and $92.8$ minutes, respectively, while for $x=13$ it is  $108.8$ minutes (Table 1). Remember that the whole point of partitioning the baselines was to cut down the computations. We also can benefit from this approach to speed up the initial calibration iterations for the telescope with a large number of baselines such as SKA. 

As a final remark, for partiting baselines of a telescope with an odd number of receivers $N$, an alternative would be: (i) first partitioning baselines for $N-1$ number of receivers, as it is already explained in this section, and (ii) assigning the remained baselines to these $N-1$ partitions.  

\subsection{Averaging of visibilities} 
The OS calibration method divides the data into sub-observations and alternates. The use of fewer data samples in each iteration is the principle cause of the speedup. So far, we have used segments of data consisting of multiple integrations in time and have considered the individual integrations as the sub-observations.  This is reasonable for the use of OS calibrations. However, for the non-OS type calibrations all of these integrations are explicitly considered to be equivalent. Therefore, one could ask if it is easier to average the data before calibration to decrease the computational cost.  \\
To answer this question, consider the case of calibrating data for a point source far away from the phase center of an observation. Based on (\ref{s3}), the visibilities of baseline $p-q$ at every sub-observation are formulated as 
\begin{equation}
{\bf v}_{pq}={\bf J}^*_{q}\otimes{\bf J}_{p}\mbox{vec}({\bf C})+{\bf n}_{pq},\label{kh2}
\end{equation}
where,
\begin{equation}
{{\bf{C}}}=e^{\frac{-2\pi j \xi}{c} (ul+vm+w(\sqrt{1-l^2-m^2}-1))}
\left[ \begin{array}{cc}
\frac{{{I}}}{2}&0\\
0&\frac{{{I}}}{2}
\end{array}\right].\label{kh22}
\end{equation} 
In (\ref{kh22}), $j^2=-1$, $\xi$ is the frequency of the observation, $c$ is the speed of light, $(l,m)$ are the source direction components corresponding to the observation phase center, $(u,v,w)$ are the geometric components of baseline $p-q$, and $I$ is the intensity of the source. \\
Since the source is far away from the phase center, $(l,m)$ in (\ref{kh22}) are large. Therefore, even very small variation of the baselines $(u,v,w)$ on different sub-observations cause huge differences in the phase terms of (\ref{kh22}). Subsequently, averaging the visibilities of (\ref{kh2}) causes de-correlation (losing amplitude) and smearing effects in the calibration residuals. 

To illustrate this, we simulate a 12 hour observation of WSRT from a very bright source with 130 Jy intensity is simulated. The source is about four degrees away from the phase center. In the center of the field we also put twenty three faint sources with intensities below 9 Jy. The Jones matrices for the faint sources are considered as identity matrices. For the bright source, they are multiplications of different linear combinations of $sin$ and $cos$ functions which are invariant on twenty five seconds time intervals. That provides time samples of size $\tau=25$ including sub-observations, from every individual second, for which the gain errors are exactly the same. White Gaussian noise is also added to the simulated data. 
\begin{figure}
\centering
\epsfig{file=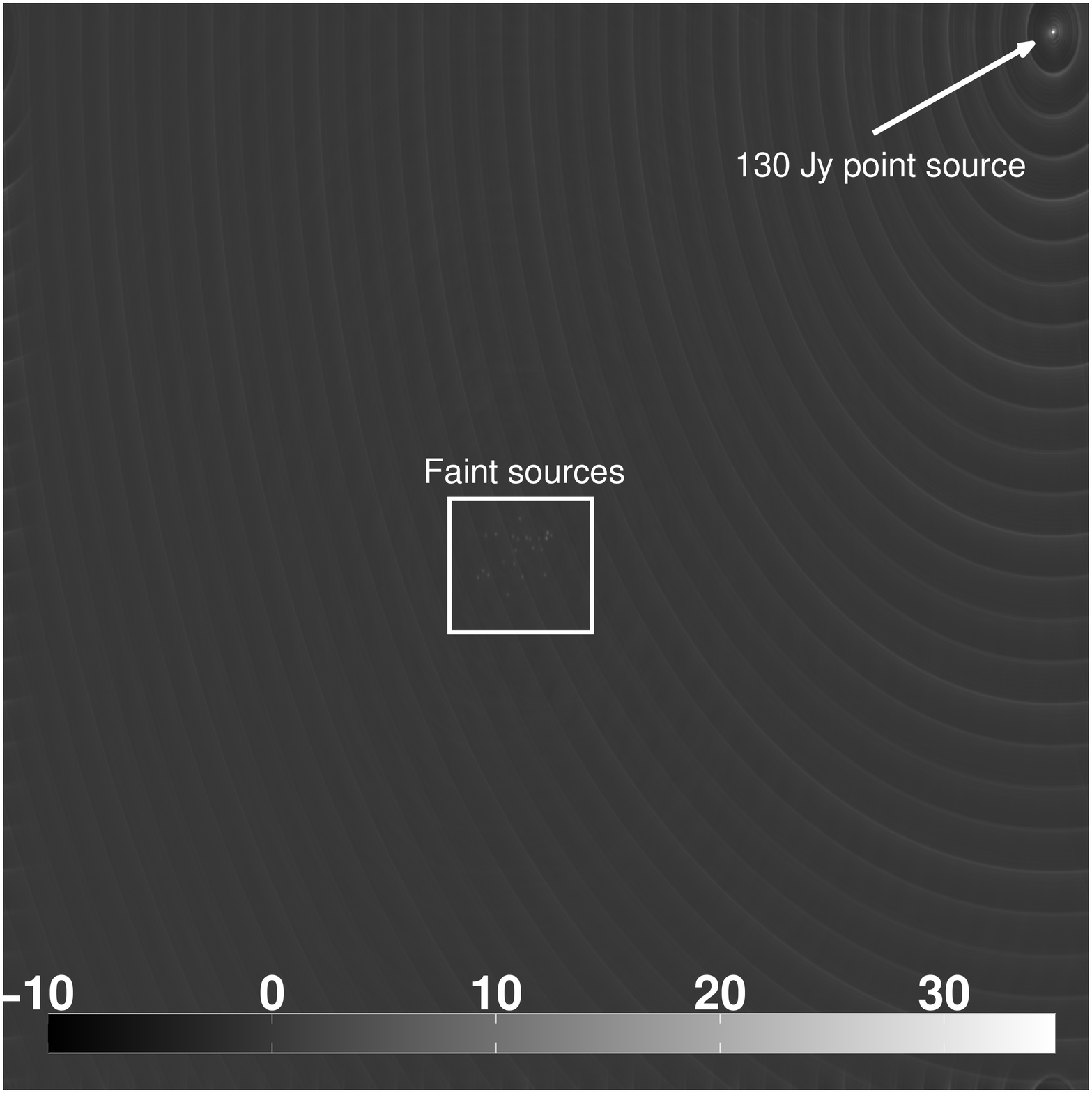, bb= -82 10 697 800,clip=,width=8.8cm,scale=0.4}
  \caption{ A 12 hour simulation of WSRT with a bright, 130 Jy source. The source is about four degrees away from the phase center. There also exist twenty three other faint sources with intensities below 9 Jy in the center of the field. White Gaussian noise is also added to the simulated data.}\label{FigureB1}
\end{figure} 

\begin{figure*}\vspace*{10cm}
\begin{center}$
\begin{array}{cc}
\multicolumn{2}{c}{\hspace{-6mm}\epsfig{file=colormap.eps, bb= -98 393 637 438, scale=0.71}}\vspace*{-9.9cm}\\
\hspace{-1mm}\epsfig{file=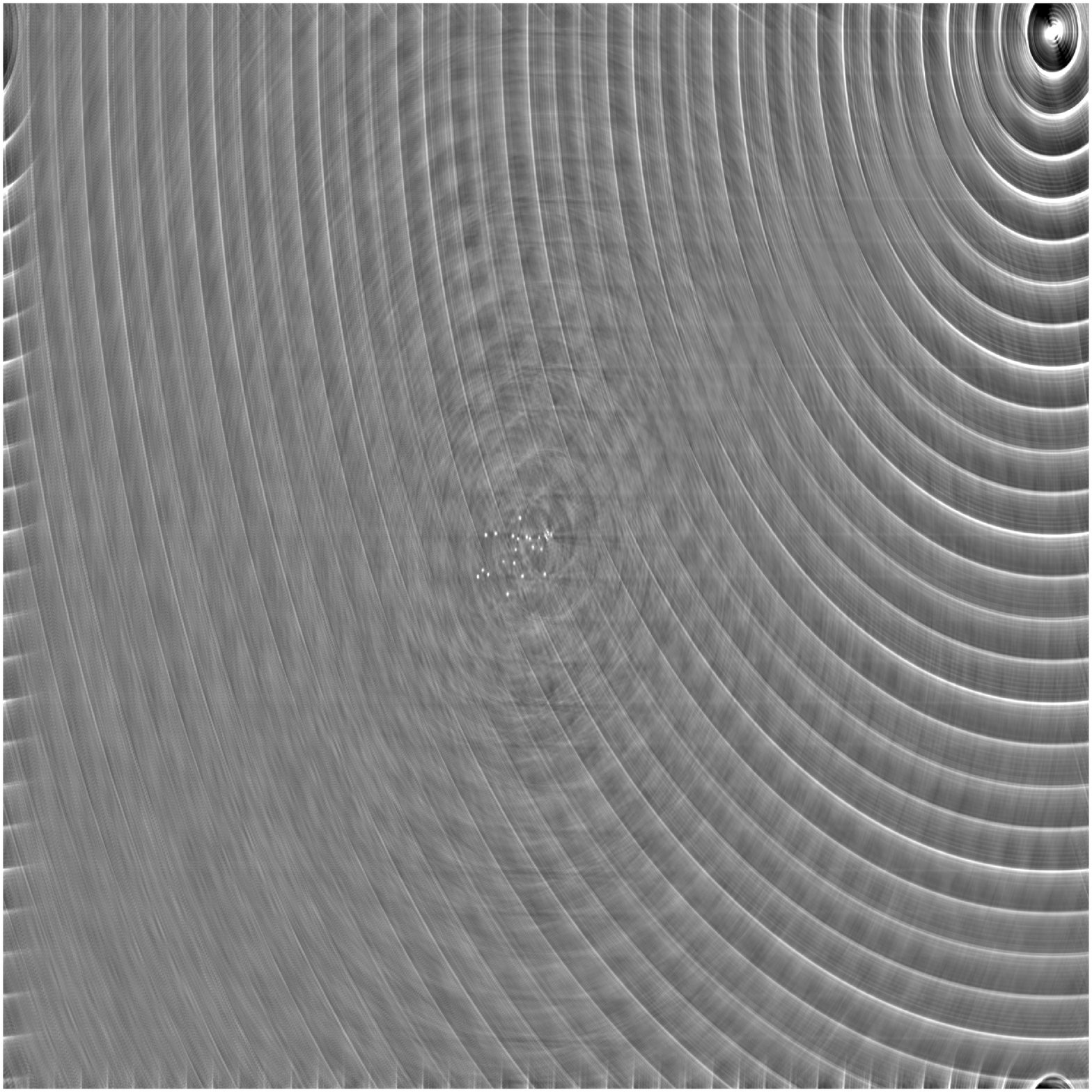, bb= -95 4 708 805,clip=,width=8.8cm,scale=0.4} &
\hspace{-3mm}\epsfig{file=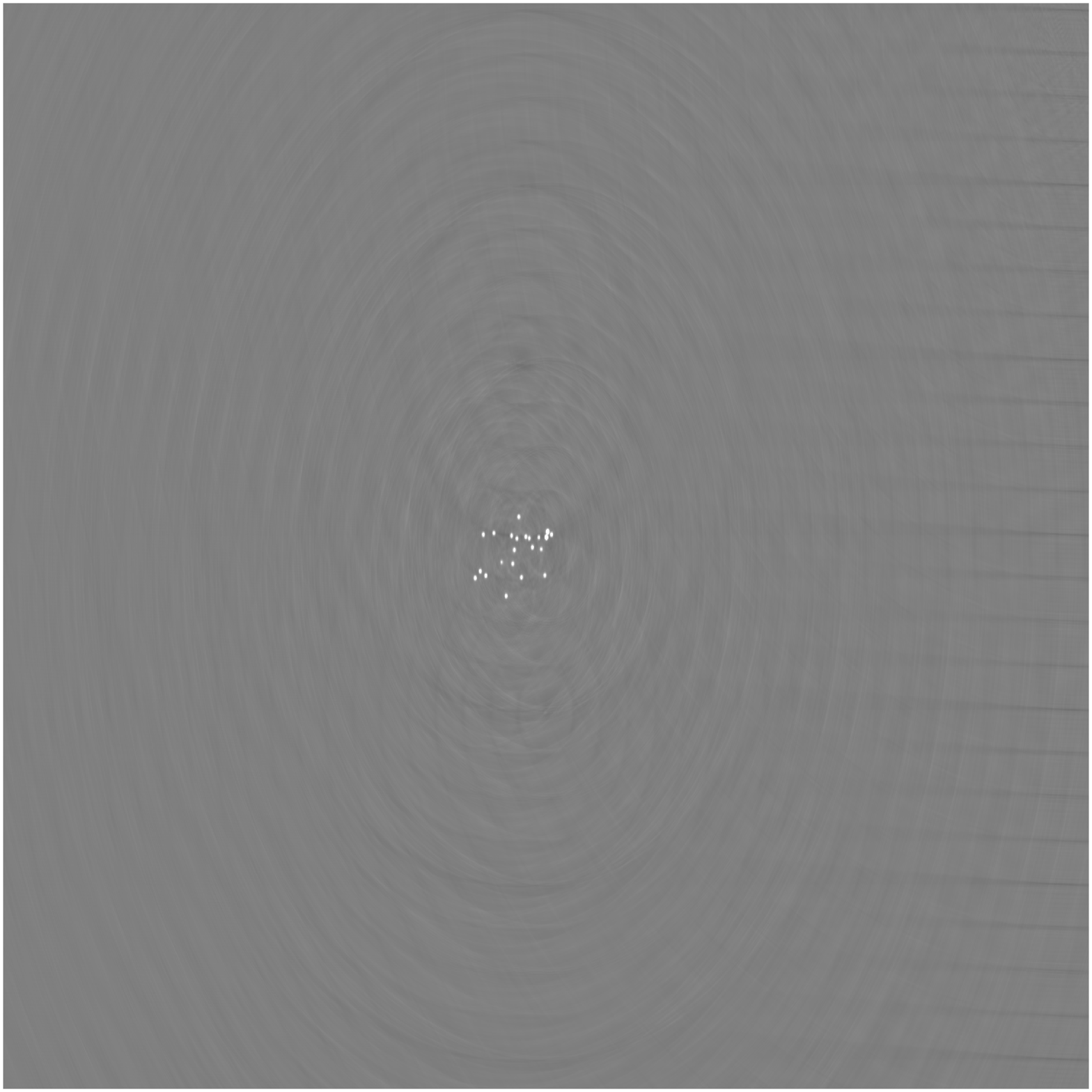, bb= -95 4 708 805,clip=,width=8.8cm,scale=0.4}\vspace*{-6mm}\\
\hspace*{3.8cm}\mbox{\color{white}\large{\bf (a)}}\quad\quad\quad\quad\quad\quad \quad\quad\quad\quad\mbox{\color{white}\large{\bf 7.63 min}}&\hspace*{3.5cm}\mbox{\color{white}\large{\bf (b)}}\quad\quad\quad\quad\quad\quad \quad\quad\quad\quad\mbox{\color{white}\large{\bf 19.43 min}}\vspace*{1cm}\\
\end{array}$
\caption{The residual images obtained by the LS and the OS-LS calibrations, utilizing $m=2$ number of OS iterations and $n=9$ number of LS iterations. The processing time is shown at the bottom right corner of every image. LS calibration on averaged visibilities (a), and OS-LS (b) calibration are applied. In (a), the bright source is highly underestimated (almost not subtracted at all) and there exist severe smearing effects. This is due to the de-correlation by averaging the visibilities. However, in (b),  for which OS-LS calibrations is applied on individual integrations, the bright source is perfectly subtracted and the other fainter sources are completely visible. That makes OS-LS calibration the method of choice, despite its longer execution time.}\label{FigureB2}
\end{center}
\vspace*{-5mm}
\end{figure*}

It is expected that traditional calibration after averaging data performs as equivalent as the OS calibration which iterates on the individual sub-observations. The reason is that the simulated corruptions in the signals on twenty five seconds time intervals are invariant. However, the results, illustrated by Fig \ref{FigureB2}, is completely the opposite. 

Fig. \ref{FigureB2} shows the residual images obtained by the LS and the OS-LS calibration, utilizing $m=2$ number of OS iterations and $n=9$ number of LS iterations. The processing time in min is shown at the bottom right corner of every image. In image (a) of Fig. \ref{FigureB2}, LS calibration is applied on averaged data obtained from $\tau=25$ time samples. In this image, the bright source is highly underestimated (almost not subtracted at all) and there exist elongated radial features. This is due to the de-correlation by averaging the visibilities. However, in image (b) of Fig. \ref{FigureB2}, for which OS-LS calibration is applied on individual integrations, the bright source is perfectly subtracted and the other fainter sources are completely visible. This proves that we can not simply apply calibration on averaged visibilities to cut down the computations and reveals the need of using the OS calibration. We have also executed LS calibration on non-averaged data sets of $\tau=25$ time samples. The resulted residual image has been exactly the same as image (b) of Fig. \ref{FigureB2}, which is generated by OS-LS calibration. The reason is that the Jones matrices on every twenty five seconds calibrated data are invariant. Therefore, the solution which is obtained by OS calibrations, using few integrations (sub-observations) within twenty five seconds, is the same as the one obtained by non-OS calibrations using all the data. However, in reality,  Jones matrices vary with time. In such a case, the result of the non-OS calibrations is always better than, or equivalent to, the one of OS calibrations. It is because finding a global solution which fits all data is generally more efficient than solving only for a part of dataset. The execution time of the LS calibration was $78.15$ min, which is indeed longer than the one of OS-LS calibration ($19.43$ min).

\section{Conclusions}
\label{Conclusions}
This paper introduces OS-LS and OS-SAGE radio interferometric calibration, as combinations of the OS method with LS and SAGE calibration techniques. We show that the OS calibration provide a significant improvement in the execution speed compared to the conventional (non-OS) calibration algorithms. The key idea is to partition the observed data into groups of sub-observations for which the gain errors are considered to be fixed. OS type calibrations solve for every group by iteratively updating the solutions for that group's sub-observations in an ordered sequence. The calibrations benefit from very fast computations and preserve almost the same quality as the one obtained by the non-OS calibrations. But, we must take in to account that their accuracy never becomes higher than the one of the non-OS calibration. Simulations show that OS calibration methods have considerable computational improvements compared to the conventional non-OS calibration methods. They also indicate that the OS-SAGE calibration provides a better quality results in a shorter time compared to the OS-LS calibration, as it is the case for the conventional SAGE and LS calibrations. In Future work, we address a novel accuracy of calibration obtained via a hybrid of non-OS and OS calibration techniques which has a computational cost almost as cheap as the one of OS calibrations.
\bibliographystyle{mn2e}
\bibliography{references}

\bsp
\label{lastpage}
 \end{document}